\documentclass[11pt]{article}
\usepackage{amsmath}
\usepackage{epsfig}

\title{Feynman Diagrams of Generalized Matrix Models
       and the Associated Manifolds in
       Dimension 4}
\author{Roberto {\titsc De Pietri}\\
Dipartimento di Fisica, Universit\`a di Parma, and \\
INFN gruppo collegato di Parma\\
Parco Area delle Scienze 7/A, I-43100 Parma, Italy\\
depietri@pr.infn.it
\and
Carlo {\titsc Petronio}\\
Dipartimento di Matematica Applicata, Universit\`a di Pisa\\
Via Bonanno Pisano 25B, I-5616 Pisa, Italy\\
petronio@dm.unipi.it}
\date{\today}

\makeatletter
\def\@begintheorem#1#2{\it \trivlist \item[\hskip \labelsep{\bf #1\ #2.}]}
\newtheorem{teo}{Theorem}[section]
\newtheorem{rem}[teo]{Remark}

\newtheorem{prop}[teo]{Proposition}
\newtheorem{que}[teo]{Question}
\newtheorem{definition}[teo]{Definition}

\def\finedim#1{{\hfill\hbox{\enspace\fbox{\ref{#1}}}}\vspace{5pt}}
\def\dim#1{\vspace{1pt}\noindent{\it Proof of} {\hspace{2pt}}\ref{#1}.}

\reversemarginpar

\newfont{\Bbb}{msbm10 scaled 1200}
\def\matR{{\hbox{\Bbb R}}}

\newfont{\Got}{eufm10 scaled 1200}
\def\permu{{\hbox{\Got S}}}

\def\calP{{\cal P}}
\def\calS{{\cal S}}
\def\calK{{\cal K}}
\def\calT{{\cal T}}

\font\titsc=cmcsc10 scaled 1200

\def\figura#1#2#3{
\begin{figure}
\centerline{\epsfig{file=#1.eps,height=#2}}
\caption{\label{#1}#3}
\end{figure}}

\def\Dir{\textsf{Dir}}
\def\Ori{\textsf{Ori}}
\def\Surf{\textsf{Surf}}
\def\Cycl{\textsf{Cycl}}
\def\lk{{\rm lk}}

\def\FG#1{\textsf{FG}_{#1}}
\def\OFG#1{\FG{#1}^{+}}
\def\GS#1{\textsf{GS}_{#1}}
\def\OGS#1{\textsf{GS}_{#1}^{+}}

\def\sgn{{\rm sgn}}

\def\idx#1{{\scriptscriptstyle (\! #1 \!)}}
\def\Balpha{{\mathbf{\alpha}}}

\begin{document}

\maketitle
\begin{abstract}
\noindent
The problem of constructing a quantum theory of gravity
has been tackled with very different strategies, most of which
relying on the interplay between ideas from physics and from
advanced mathematics. On the mathematical side, a central r\^ole
is played by combinatorial topology, often used to recover the
space-time manifold from the other structures involved. An
extremely attractive possibility is that of encoding all possible
space-times as specific Feynman diagrams of a suitable field
theory. In this work we analyze how exactly one can associate
combinatorial 4-manifolds to the Feynman diagrams of certain
tensor theories.

\vspace{.2cm}
\vspace{.1cm}\vspace{.2cm}
\end{abstract}

\section{Introduction}
We describe in this paper precise conditions which allow to
associate a four-dimensional manifold to a Feynman diagram of a
rank-four tensor theory. The question originates from the attempt
of formulating a quantum theory of gravity \cite{JM} in terms of a
generalization of the matrix model formulation of two-dimensional
quantum gravity \cite{2d}. We try to keep our discussion as much
as possible independent of dimension, but we only give precise
conditions in dimension two, three and four. The result was
actually previously known in dimension two and three, but the
fact that we can give a unified treatment allows us to clarify
better the specific conditions which we exhibit in dimension four.

To give an $n$-dimensional theory of Euclidean quantum gravity one
needs to define a path integral of the following form over equivalence
classes of metrics:
\begin{equation}
  \label{eq:ZQG}
  Z[\Lambda,G] = \sum_{M\in {\rm Top}_n}\ \ \int\limits_{{\rm Riem}(M)/{\rm Diff}(M)}
          \mathcal{D}[g] \;{\rm e}^{-S_g[\Lambda,G]}
  \; .
\end{equation}
Here ${\rm Riem}(M)/{\rm Diff}(M)$ is the space of
Riemannian metrics over $M$ modulo diffeomorphism,
the weight factor $S_g[\Lambda,G]$ is the
standard Einstein-Hilbert action
\begin{equation}
\nonumber 
  S_g[\Lambda,G]= \int_M \!\! \sqrt{g} d^n\!x  \left(
                   - \frac{1}{16\pi G}  R[g]
                 + \Lambda \right)
  \; ,
\end{equation}
and a sum over all possible $n$-dimensional topologies has also been included.
Formal expressions like (\ref{eq:ZQG}) have demonstrated extremely
powerful heuristic tools in theoretical physics but, as a matter of fact,
they generally lack a proper definition. To actually give a definition one
typically needs to be more specific and restrictive about the class
${\rm Top}_n$ of manifolds under consideration, and about the space
${\rm Riem}(M)/{\rm Diff}(M)$ of metrics. A possible
strategy is to consider triangulated manifolds and
to approximate path integration over inequivalent
Riemannian structures with summation over inequivalent
triangulations of the manifold. Here a triangulation
$\calT$ is considered to define the {\em singular}
Riemannian structure in which all $n$-simplices are flat and have edges
of a given length $a$. The action takes the form
\begin{eqnarray}
\label{eq:action:for:triangulation}
  S(\Lambda,G;\calT) &=& k_n \nu_n(\calT) - h_{n-2} \nu_{n-2}(\calT)
\end{eqnarray}
where $\nu_i(\calT)$ is the number of $i$-simplices in $\calT$ and
$k_i$, $h_i$ are given as follows in terms of the
volume $\mathrm{vol}(\sigma^i)$ of the flat $i$-simplex with edges
having length $a$:
\begin{equation}
\nonumber 
  \begin{array}{lcl}
  k_{n}   &=&  \Lambda\cdot \mathrm{vol}(\sigma^n)
           + \frac{n(n+1)}{2}\frac{\arccos(1/n)}{16\pi G}
             \mathrm{vol}(\sigma^{n-2}),  \\
  h_{n-2} &=& \frac{1}{8G} \mathrm{vol}(\sigma^{n-2}).
  \end{array}
\end{equation}
This approach is known as `Dynamical Triangulation'
(see~\cite{Carfora} for discussion and references)
and the partition function $Z[\Lambda,G]$ is given by
\begin{equation} \label{Z:DT}
Z_{{\rm DT}}[\Lambda,G] = \sum_{M\in {\rm Top}_n}\
      \sum_{\calT\in {\rm Tria}(M)}
      \;{\rm e}^{-k_n \nu_n(\calT) + h_{n-2} \nu_{n-2}(\calT)}.
\end{equation}
Here ${\rm Tria}(M)$ denotes the set of triangulations, up to
combinatorial equivalence, of a manifold $M$, and the partition
function is perfectly analogous to that of~(\ref{eq:ZQG}).  Note that
the sum over triangulations together with the sum over topologies
corresponds to the sum over all $n$-dimensional simplicial complexes
which are manifolds.  (As usual we consider also ``singular''
triangulations, {\em i.e.}~we allow self-adjacencies and multiple
adjacencies.) Note also that if in (\ref{eq:ZQG}) one wants to
consider the action on ${\rm Riem}(M)$ of only those diffeomorphisms
which are isotopic to the identity, then the discretized analogue
again has the form of (\ref{Z:DT}), except that each triangulation
$\calT$ should now be weighted according to the number of non-isotopic
ways it can be realized in the corresponding manifold.

Dynamical Triangulations have been intensively investigated in two,
three and four dimensions. It is of particular interest that in two
dimensions the theory may be reformulated in terms of a matrix
model~\cite{2d}. In the perturbative approach to this theory, the
resulting Feynman diagrams have vertices which correspond to
two-simplices, and propagators which correspond to edge-pairings, so a
diagram leads to a surface obtained by glueing triangles.  In
extending the triangulation strategy to arbitrary dimension one is
brought to the search for theories having Feynman diagrams in which
vertices can be identified with $n$-simplices, and propagators with
glueings of codimension-1 faces. If this happens then each Feynman
diagram gives an $n$-dimensional simplicial complex.

Generalized matrix models with Feynman diagrams corresponding to
simplicial complexes were discussed by Sasakura \cite{TM3}, Gross
\cite{Gross}, Boulatov \cite{Boulatov:1992}, Ooguri
\cite{Ooguri:1992b}, and more recently by De
Pietri-Freidel-Krasnov-Rovelli\cite{DePietri:2000} and others.
However, in all these works a complete discussion of the
topological properties of the resulting simplicial complexes is
missing, and the results of the present paper allow to fill this
gap. More precisely, it is the aim of the present paper to
introduce a specific $n$-tensor model whose Feynman diagrams can
be identified with $n$-dimensional simplicial complexes, and to
explicitly analyze the topological properties of these spaces.
After a brief discussion of the two- and three-dimensional cases
we address the four-dimensional problem. As a main result we
provide an explicit criterion to decide whether the simplicial
complex associated to a given Feynman diagram is a manifold or
not. When the criterion is fulfilled the value of a Feynman
diagram corresponds to the value of the discretized
Einstein-Hilbert action of~(\ref{eq:action:for:triangulation}) on
the associated triangulation. In connection with the papers
mentioned above, it is worth remarking here that the weight
factor of model \cite{Boulatov:1992} corresponds to the three
dimensional Ponzano-Regge \cite{Ponzano-Turaev-Viro} model and the
one of model \cite{DePietri:2000} to the Barrett-Crane
\cite{Barrett:1998} four dimensional relativistic spin model,
while in our model each Feynman graph is weighted by the
simplicial action (\ref{eq:action:for:triangulation}). Moreover,
as consequence of Remark \ref{circuit:rem}, all these models can
be seen as Spin Foam \cite{SpinFoam} models.

\section{Generalized matrix models and fat graphs}

We review in this section the matrix model whose Feynman diagrams lead
to surfaces, then we introduce more complicated models which
eventually will give higher-dimensional simplicial complexes.
We set up a useful graphical encoding of the Feynman diagrams, which
will help us in describing how exactly a diagram leads to a simplicial
complex.

\paragraph{Two-dimensional quantum gravity as a matrix model}
The partition function of the matrix model corresponding to two-dimensional
quantum gravity is
\begin{equation}
  \label{eq:Zmatrixmodel}
  Z[N,\lambda] = \int [d\phi] \;{\rm e}^{-\frac{1}{2} \mathrm{Tr}[\phi^2]
                         +\frac{\lambda}{3} \mathrm{Tr}[\phi^3]}
  ,\qquad [d\phi]=\frac{1}{\mathcal{N}}
                  \prod_{\alpha\leq\beta} d\Re{\phi_{\alpha\beta}}
                  \prod_{\alpha<\beta} d\Im{\phi_{\alpha\beta}}
\end{equation}
where the configuration variable is a Hermitian
$N\times N$ matrix $\phi=(\phi_{\alpha\beta})$, and
$\mathcal{N}$ is a normalization constant chosen
such that $Z[N,0]=1$.
We note that the integral defining $Z[N,\lambda]$
is divergent for real $\lambda$. This means that some
procedure must be given to properly define~(\ref{eq:Zmatrixmodel}).
For the purpose of this work we will consider $Z[N,\lambda]$ as a formal
power series in $\lambda$ and we will view~(\ref{eq:Zmatrixmodel})
as a formal definition of the coefficients
$Z^{(\!k\!)}[N]$ such that
$Z[N,\lambda]= \sum_k Z^{(\!k\!)}[N] \lambda^k/k!$, where
$Z^{(\!k\!)}[N]$ corresponds to
$d^k Z[N,\lambda]/d\lambda^k |_{\lambda=0}$, and is
computed by interchanging integration and derivation.
In the standard field-theoretical language we say that we
evaluate the integral of~(\ref{eq:Zmatrixmodel}) as the Feynman
graph expansion (formal power series in $\lambda$) generated by
\begin{eqnarray}
\nonumber 
&&  Z[N,\lambda] = \left.
         \exp\left[
            \frac{\lambda}{3}
            V^{\alpha_1\alpha_2;\beta_1\beta_2;\gamma_1\gamma_2}
            \frac{\delta}{\delta J^{\alpha_1\alpha_2}}
            \frac{\delta}{\delta J^{\beta_1\beta_2}}
            \frac{\delta}{\delta J^{\gamma_1\gamma_2}}
           \right] Z^{\scriptscriptstyle (0)}[N;J]
        \right|_{J=0}
\\ \nonumber 
&&  Z^{\scriptscriptstyle (0)}[N;J]
       = \int [d\phi] \;{\rm e}^{-\frac{1}{2} \mathrm{Tr}[\phi^2]
                         + \phi_{\alpha\beta} J^{\alpha\beta} }
       = {\rm const} \cdot \exp\left[\frac{1}{2} J^{\alpha\beta}
                 G_{\alpha\beta;\gamma\delta} J^{\gamma\delta}
               \right]
  , \qquad
\end{eqnarray}
where the Einstein convention of sum over repeated
covariant-contravariant indices is assumed,
$G_{\alpha_1\alpha_2;\beta_1\beta_2}
= g_{\alpha_1\beta_2}\cdot g_{\alpha_2\beta_1}$,
$V^{\alpha_1\alpha_2;\beta_1\beta_2;\gamma_1\gamma_2}
= g^{\alpha_2\beta_1}\cdot g^{\beta_2\gamma_1}
g^{\gamma_2\alpha_1}$,
$g_{\alpha\beta}=g^{\alpha\beta}=\delta^\alpha_\beta$,
where $\delta$ is the Kronecker symbol. The function
$Z^{\scriptscriptstyle (0)}[N;J]$ thus defined is the free partition function
in presence of the source $J$.

It is now convenient to use a graphical representation of
tensor expressions containing sums over dummy indices.
We will represent the metric and $\delta$ as:
\begin{equation}\label{eq:GR}
  g_{\alpha_1\alpha_2}
  = \begin{array}{c}\setlength{\unitlength}{1 pt}
    \begin{picture}(20,30)
    \put( 0,20){\makebox(10,10){$\scriptstyle \alpha_1$}}
    \put(10,20){\makebox(10,10){$\scriptstyle \alpha_2$}}
    \put(10,20){\oval(10,20)[b]}
    \end{picture}\end{array}
  ,\qquad
  g^{\beta_1\beta_2}
  =\begin{array}{c}\setlength{\unitlength}{1 pt}
    \begin{picture}(20,30)
    \put( 0, 0){\makebox(10,10){$\scriptstyle \beta_1$}}
    \put(10, 0){\makebox(10,10){$\scriptstyle \beta_2$}}
    \put(10,10){\oval(10,20)[t]}
    \end{picture}\end{array}
  ,\qquad
  \delta^{\alpha}_{\beta}
  =\begin{array}{c}\setlength{\unitlength}{1 pt}
    \begin{picture}(10,30)
    \put( 0, 0){\makebox(10,10){$\scriptstyle \beta$}}
    \put( 0,20){\makebox(10,10){$\scriptstyle \alpha$}}
    \put( 5,10){\line(0,1){10}}
    \end{picture}\end{array}
\end{equation}
Moreover, we will use the following symbols
\begin{equation}\label{eq:GRbis}
T^{\alpha_1\ldots\alpha_i}_{\beta_1\ldots\beta_j}
= \begin{array}{c}\setlength{\unitlength}{1 pt}
  \begin{picture}(50,50)
    \put( 0,37){\makebox(10,10){$\scriptstyle \alpha_1$}}
    \put(20,37){\makebox(10,10){$\ldots$}}
    \put(40,37){\makebox(10,10){$\scriptstyle \alpha_i$}}
    \put( 5,35){\line( 0,-1){ 5}}
    \put(45,35){\line( 0,-1){ 5}}
    \put( 0,20){\framebox(50,10){$T$}}
    \put(10,20){\line( 0,-1){ 5}}
    \put(40,20){\line( 0,-1){ 5}}
    \put( 5, 3){\makebox(10,10){$\scriptstyle \beta_1$}}
    \put(20, 3){\makebox(10,10){$\ldots$}}
    \put(35, 3){\makebox(10,10){$\scriptstyle \beta_j$}}
\end{picture}\end{array}
\qquad,\qquad
\prod_{i=1}^n \delta^{\alpha_i}_{\beta_{\sigma(i)}}
= \begin{array}{c}\setlength{\unitlength}{1 pt}
  \begin{picture}(50,50)
    \put( 0,37){\makebox(10,10){$\scriptstyle \alpha_1$}}
    \put(10,37){\makebox(10,10){$\scriptstyle \alpha_2$}}
    \put(25,37){\makebox(10,10){$\ldots$}}
    \put(40,37){\makebox(10,10){$\scriptstyle \alpha_n$}}
    \put( 5,35){\line( 0,-1){ 5}}
    \put(15,35){\line( 0,-1){ 5}}
    \put(45,35){\line( 0,-1){ 5}}
    \put( 0,20){\framebox(50,10){$\sigma\; {\scriptstyle\uparrow}$}}
    \put( 5,20){\line( 0,-1){ 5}}
    \put(15,20){\line( 0,-1){ 5}}
    \put(45,20){\line( 0,-1){ 5}}
    \put( 0, 3){\makebox(10,10){$\scriptstyle \beta_1$}}
    \put(10, 3){\makebox(10,10){$\scriptstyle \beta_2$}}
    \put(25, 3){\makebox(10,10){$\ldots$}}
    \put(40, 3){\makebox(10,10){$\scriptstyle \beta_n$}}
\end{picture}\end{array}
\end{equation}
to represent a generic tensor with $i$ contravariant
and $j$ covariant indices, and a permutation
$\sigma\in\permu_n$, respectively.
When two strands are connected by a line we assume that the strands carry
the same dummy index and that summation over this index is taken.
Using these conventions, the  propagator and the vertex are represented by
the diagrams:
\begin{eqnarray*}
V^{\alpha_1\alpha_2;\beta_1\beta_2;\gamma_1\gamma_2}
&\Rightarrow&
   \begin{array}{c}\setlength{\unitlength}{1 pt}
   \begin{picture}(100,25)
            \put(45,15){\oval(90,30)[b]}
            \put(25,15){\oval(30,24)[b]}
            \put(65,15){\oval(30,24)[b]}
            \put( 0,17){${\scriptstyle \alpha_1\alpha_2}$}
            \put(40,17){${\scriptstyle \beta_1\beta_2}$}
            \put(80,17){${\scriptstyle \gamma_1\gamma_2}$}
   \end{picture}\end{array}
\\
G_{\alpha_1\alpha_2;\beta_1\beta_2}
&\Rightarrow&
   \begin{array}{c}\setlength{\unitlength}{1 pt}
   \begin{picture}(80,25)
            \put(35,10){\oval(70,26)[t]}
            \put(35,10){\oval(50,20)[t]}
            \put( 0, 0){${\scriptstyle \alpha_1\alpha_2}$}
            \put(60, 0){${\scriptstyle \beta_1\beta_2}$}
   \end{picture}\end{array}
\end{eqnarray*}
A direct inspection shows that, up to order $\lambda^2$, the formal power
series $Z[N,\lambda]$ is given by
\begin{eqnarray*}
  Z[N,\lambda] & = & 1 + \lambda^2 \left( \frac{1}{6} [[D1]]
           + \frac{1}{2}[[D2]] + \frac{1}{6} [[D3]]
           \right) + \ldots \\
               & = & 1 + \lambda^2 \left( \frac{1}{6} N^3
           + \frac{1}{2} N^3 + \frac{1}{6} N \right) + \ldots\ .
\end{eqnarray*}
where $[[D1]]$, $[[D2]]$ and $[[D3]]$ are the evaluations of the
Feynman diagrams of Fig.~\ref{exTM2}, carried out using the
correspondence just discussed between diagrams and tensor
expressions. Note that, if we associate a triangle to each vertex
and an edge-glueing to each propagator as described in
Fig.~\ref{exTM2}, then the two diagrams corresponding to the
sphere evaluate to $N^3$, while the one corresponding to the
torus evaluates to $N$.

\figura{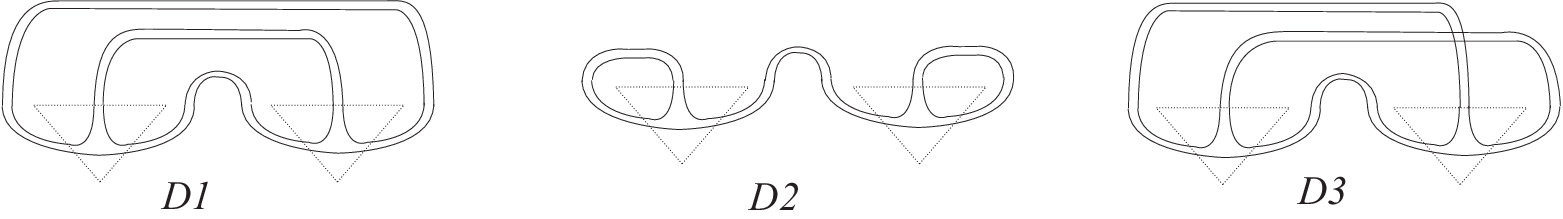}{2cm}{The three inequivalent Feynman diagrams of the
   two-dimensional matrix model~\protect{(\ref{eq:Zmatrixmodel})}
   at order \protect{$\lambda^2$}. The association of triangles to
   vertices is explicitly shown. Interpreting each propagator as a gluing
   instruction between two edges of triangles, it is easy to see that \protect{$D1$}
   and \protect{$D2$} correspond to different triangulation of the sphere,
   while  \protect{$D3$} corresponds to a triangulation of the torus.}

\paragraph{Higher-rank generalized matrix models}

It is possible to consider higher-dimensional
extensions of the matrix model~(\ref{eq:Zmatrixmodel}),
using higher-rank tensors. The natural generalization of
the matrix model partition function is achieved by considering as
configuration variable an $n$-tensor ${\phi_{\alpha_{1}\ldots\alpha_{n}}}$,
where each $\alpha_i$ varies between $1$ and $N$,
having the following symmetry:
\begin{equation} \label{GMM:real}
  \phi_{\alpha_{\tau(1)} \ldots \alpha_{\tau(n)} }
  =    \Re [ {\phi_{\alpha_{1}\ldots\alpha_{n}}} ]
   + i\cdot \sgn(\tau)\cdot \Im [{\phi_{\alpha_{1}\ldots\alpha_{n}}} ]
\end{equation}
where $\tau\in\permu_n$ and $\sgn(\tau)$ is the signature
(also called parity) of $\tau$. Using multi-indices
$\Balpha=(\alpha_1\ldots\alpha_n)$ we consider
the following partition function for the $n$-tensor model
\begin{equation}\label{Z:ntensor}
Z_n[N,\lambda] = \int [d\phi] \;\exp \left[
          {-\frac{1}{2} \sum_\Balpha |\phi_\Balpha|^2
          +\frac{\lambda}{n+1}
           \sum_{\Balpha^\idx{0}\ldots \Balpha^\idx{n}}
              V^{\Balpha^\idx{0}\ldots \Balpha^\idx{n}}
              \cdot\phi_{\Balpha^\idx{0}}\cdot\ldots\cdot \phi_{\Balpha^\idx{n}}
          }\right]
\end{equation}
where $V^{\Balpha^\idx{0}\ldots \Balpha^\idx{n}}$ is the
vertex function which will be defined in (\ref{EQ:genVertex}) below
(we do not define it here because its form is not needed and because the
general definition will involve notations introduced later. A definition
for the three-tensor and four-tensor model will be given before the general one
in (\ref{eq:Zmatrix3}) and (\ref{eq:Zmatrix4}), respectively).

As in the 2-dimensional case, the expansion of $Z_n[N,\lambda]$ is
obtained by introducing another function $Z_n^{\scriptscriptstyle (0)}[N;J]$
such that
\begin{equation}\label{free:non-free:rel}
Z_n[N,\lambda] = \left. \exp\left[
\frac{\lambda}{n+1} V^{\Balpha^\idx{0}\ldots \Balpha^\idx{n}}
\frac{\delta}{\delta J^{\Balpha^\idx{0}}}\cdots
\frac{\delta}{\delta J^{\Balpha^\idx{n}}}\right]
Z_n^{\scriptscriptstyle (0)}[N;J]
\right|_{J=0}.
\end{equation}

\begin{prop}\label{GM:rel} The free partition function in presence of
source of the generalized matrix models (\ref{Z:ntensor}) whose
fundamental field  fulfills the symmetry requirements (\ref{GMM:real}), defined as
$$  Z_n^{\scriptscriptstyle (0)}[N;J]
   = \int [d\phi]
   \exp\Bigg[ -\frac{1}{2} \sum_{\alpha_1,\ldots,\alpha_n}
                   |\phi_{\alpha_1\ldots\alpha_n}|^2
             + \sum_{\alpha_1,\ldots,\alpha_n}
               \phi_{\alpha_1\ldots\alpha_n} J^{\alpha_1\ldots\alpha_n}
            \Bigg]$$
is given by

\begin{equation}\label{free:part:form}
  Z_n^{\scriptscriptstyle (0)}[N;J]
   = \exp\left[\frac{1}{2}
                 J^{\alpha_1\ldots\alpha_n}
                 G_{\alpha_1\ldots\alpha_n;\beta_1\ldots\beta_n}
                 J^{\beta_1\ldots\beta_n}
               \right],
\end{equation}
where the propagator is defined as
\begin{equation} \label{eq:GMMprop}
G_{\alpha_1\ldots\alpha_n;\beta_1\ldots\beta_n}
   = \frac{2}{n!}
        \sum_{\begin{array}{c} \scriptstyle \tau\in\permu_n \\[-1mm]
                               \scriptstyle \sgn(\tau)=-1 \end{array}}
        G_{\alpha_1\ldots\alpha_n;\beta_1\ldots\beta_n}^\idx{\tau},
\end{equation}
and $G_{\alpha_1\ldots\alpha_n;\beta_1\ldots\beta_n}^\idx{\tau}
   = \delta_{\alpha_{\tau(1)}\beta_1} \ldots \delta_{\alpha_{\tau(n)}\beta_n}$.
Moreover, if integration is restricted to real $\phi$,
the partition function has the same form (\ref{free:part:form})
but the propagator is given by
\begin{equation} \label{eq:GMMprop2}
G_{\alpha_1\ldots\alpha_n;\beta_1\ldots\beta_n}
   = \frac{1}{n!}
        \sum_{\tau\in\permu_n}
        G_{\alpha_1\ldots\alpha_n;\beta_1\ldots\beta_n}^\idx{\tau}
\end{equation}
\end{prop}

\dim{GM:rel} The measure $[d\phi]$, as usual, is the product of
the Lebesgue measures over all the independent components of $\phi$
multiplied by a normalization constant $\mathcal{N}$ such that
$Z_n^{\scriptscriptstyle (0)}[N;0]=1$.

The symmetry requirement~(\ref{GMM:real}) implies that the
free partition function in presence of source can be written as follows
just in terms of the completely
symmetrized $J^{\beta_1\ldots\beta_n}_S$ and the
completely anti-symmetrized $J^{\beta_1\ldots\beta_n}_A$ parts of
the source $J^{\beta_1\ldots\beta_n}$:
\begin{eqnarray*}
& & Z_n^{\scriptscriptstyle (0)}[N;J] \\
& & =\ \ \frac{1}{\mathcal{N}} \int \!\!
  \prod_{\alpha_1\leq\ldots\leq\alpha_n}
  \!\! d\Re{\phi_{\alpha_1\ldots\alpha_n}}
     \exp\Bigg[ -\frac{1}{2} \!\sum_{\alpha_1,\ldots,\alpha_n}
                   |\Re\phi_{\alpha_1\ldots\alpha_n}|^2
             + \!\sum_{\alpha_1,\ldots,\alpha_n}
               \Re\phi_{\alpha_1\ldots\alpha_n} J_S^{\alpha_1\ldots\alpha_n}
            \Bigg] \times
\\
& & \ \ \ \ \qquad \times \prod_{\alpha_1<\ldots<\alpha_n}
   \!\!d\Im{\phi_{\alpha_1\ldots\alpha_n}}
     \exp\Bigg[ -\frac{1}{2} \!\sum_{\alpha_1,\ldots,\alpha_n}
                   |\Im\phi_{\alpha_1\ldots\alpha_n}|^2
             + i \!\sum_{\alpha_1,\ldots,\alpha_n}
               \Im\phi_{\alpha_1\ldots\alpha_n} J_A^{\alpha_1\ldots\alpha_n}
            \Bigg].
\end{eqnarray*}
Gaussian integration with respect to the variables
$\Re\phi_{\alpha_1\ldots\alpha_n}$ and
$\Im\phi_{\alpha_1\ldots\alpha_n}$ can be easily performed.
The result is indeed:
$$
  Z_n^{\scriptscriptstyle (0)}[N;J] =\frac{k}{\mathcal{N}}
  \exp\Bigg[ \frac{1}{2} \sum_{\alpha_1,\ldots,\alpha_n}
                   (J_S^{\alpha_1\ldots\alpha_n})^2
           - \frac{1}{2} \sum_{\alpha_1,\ldots,\alpha_n}
                   (J_A^{\alpha_1\ldots\alpha_n})^2
            \Bigg]
$$
where $k=k(n,N)$ is a constant which takes into account the number
Gaussian integrations which have been performed. It is now easy
to check that, fixing $\mathcal{N}=k$, the first part of the
proposition holds. Now, if the variable $\phi$ is restricted to be
real, {\em i.e.}~integration with respect to the
$\Re\phi_{\alpha_1\ldots\alpha_n}$'s is only considered, then the
term depending on $J_A^{\alpha_1\ldots\alpha_n}$ does not appear
and this proves second part of the proposition. (Note that
restricting to real $\phi$ one has to make a different choice of
$\mathcal{N}$.)\finedim{GM:rel}

Using Proposition~\ref{GM:rel}, we can now rewrite the partition
function of the $n$-tensor model in terms of its Feynman diagram
expansion. In fact, combining (\ref{free:non-free:rel}) and
(\ref{free:part:form}), it is possible to express $Z_n[N,\lambda]$
as the formal power series
\begin{equation} \label{feyn:ntensor}
\begin{split}
Z_n[N,\lambda] = \sum_k &
        \frac{1}{k!} \frac{\lambda^k}{(n+1)^k}
        V^{\Balpha^\idx{0}\ldots \Balpha^\idx{n}}\cdot \ldots\cdot
        V^{\Balpha^\idx{kn+k-n-1}
           \ldots \Balpha^\idx{kn+k-1}}
        \times\\
&\!\!\!\!\times
        \frac{(1/2)^{k(n+1)/2}}{(k(n+1)/2)!}
        \sum_{\sigma\in\permu_{\{0,\ldots,kn+k-1\}}} \!\!\!\!
        G_{\Balpha^\idx{\sigma(0)}\Balpha^\idx{\sigma(1)}}
        \cdot\ldots\cdot
        G_{\Balpha^\idx{\sigma(kn+k-2)}
           \Balpha^\idx{\sigma(kn+k-1)}}
\end{split}
\end{equation}
where the sum over $k$ is restricted to all positive $k$'s such that
$k(n+1)$ is even. The explicit form of the propagator
$G_{\Balpha^\idx{1}\Balpha^\idx{2}}$ is given by (\ref{eq:GMMprop}),
whereas the vertex has not been defined yet.
Note that restricting to real $\phi$
we would have obtained the same expression with the propagator
of (\ref{eq:GMMprop2}).

\paragraph{The three-tensor and four-tensor generalized matrix model}
Since our main interest is to analyze models related to quantum
gravity in three and four dimensions, we provide now the explicit
definition of the partition function in these two cases:
\begin{eqnarray}
  \label{eq:Zmatrix3}
  Z_3[N,\lambda] &=& \int [d\phi]
   \exp\Bigg[ -\frac{1}{2} \sum_{\alpha_1,\alpha_2,\alpha_3}
                    |\phi_{\alpha_1\alpha_2\alpha_3}|^2
\\ \nonumber && \qquad\qquad
      +\frac{\lambda}{4}
         \sum_{\alpha_1,\ldots,\alpha_6}
         \phi_{\alpha_1\alpha_2\alpha_3} \phi_{\alpha_4\alpha_5\alpha_3}
         \phi_{\alpha_4\alpha_2\alpha_6} \phi_{\alpha_1\alpha_5\alpha_6}
    \Bigg]
\\  \label{eq:Zmatrix4}
  Z_4[N,\lambda] &=& \int [d\phi]
   \exp\Bigg[ -\frac{1}{2} \sum_{\alpha_1,\ldots,\alpha_4}
                   |\phi_{\alpha_1\alpha_2\alpha_3\alpha_4}|^2
\\ \nonumber && \qquad\qquad
      +\frac{\lambda}{5}
         \sum_{\alpha_1,\ldots,\alpha_{10}}
         \phi_{\alpha_1\alpha_2\alpha_3\alpha_4}
         \phi_{\alpha_4\alpha_5\alpha_6\alpha_7}
         \phi_{\alpha_7\alpha_3\alpha_8\alpha_9}
         \phi_{\alpha_9\alpha_6\alpha_2\alpha_{10}}
         \phi_{\alpha_{10}\alpha_8\alpha_5\alpha_1}
     \Bigg]
\end{eqnarray}

Using the graphic presentation defined by (\ref{eq:GR})
and (\ref{eq:GRbis}), the Feynman rules of the two models
(\ref{eq:Zmatrix3}) and (\ref{eq:Zmatrix4}) are illustrated respectively
in Fig.~\ref{TM3dim} and in Fig.~\ref{TM4dim}.

 \figura{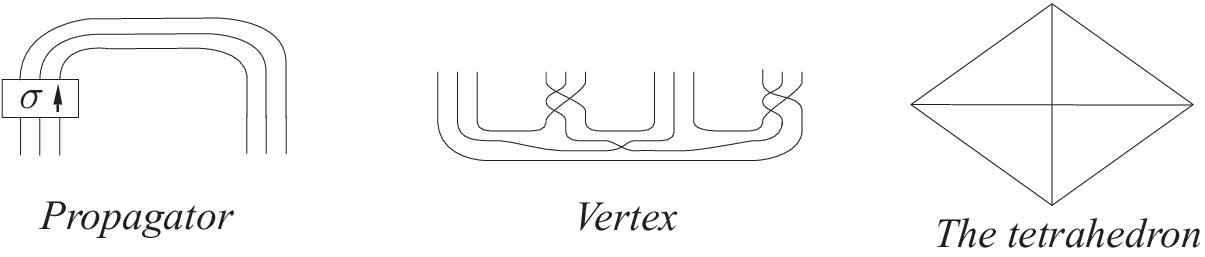}{2cm}{Feynman rules of the 3-tensor generalized matrix
 model. The analogy of the vertex diagram with the tetrahedron
 is explicitly shown.}

\begin{rem}\label{weird:3D:choice}
{\em The choice of indices in the last term of (\ref{eq:Zmatrix3})
may appear to be a weird one, since it leads to the vertex picture of
Fig.~\ref{TM3dim},
in which two of the ends of the vertex have a seemingly unnatural torsion.
An explanation of this choice will be given in Remark~\ref{all:induced:ori:rem}.}
\end{rem}

\figura{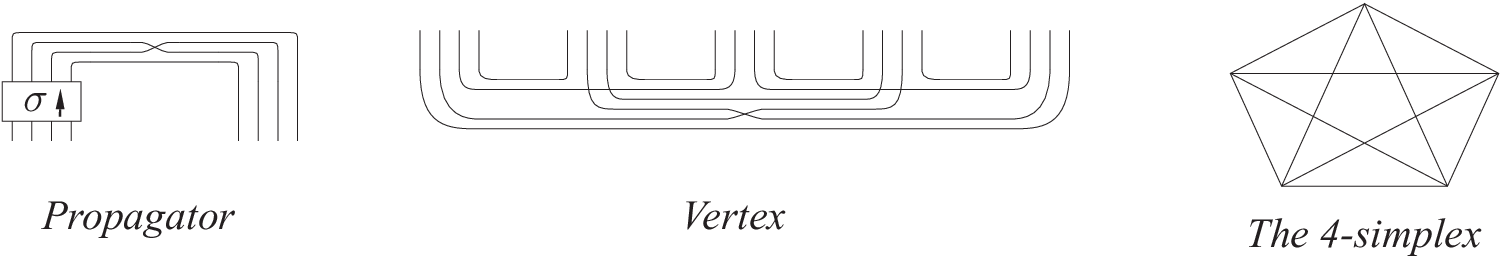}{2cm}{Feynman rules of the 4-tensor generalized matrix
model. The analogy of the vertex diagram with the 4-simplex
is explicitly shown.}

\paragraph{Encoding of Feynman diagrams: fat graphs}

We introduce in this paragraph labeled graphs
which  conveniently encode the Feynman diagrams arising from the
models discussed above.

\begin{definition}{\em  A {\em fat $n$-graph} is a connected $(n+1)$-valent
graph with the following additional structures: all the edges are oriented and
labeled by a permutation in $\permu_n$, and at each vertex the $n+1$ initial
portions of edge emanating from the vertex are numbered by $0,1,\dots,n$.
We will denote by $\FG{n}$ the set of all fat $n$-graphs up to
homeomorphisms which preserve all structures.}
\end{definition}

A practical way of describing a fat $n$-graph,
which we will always use, is as follow. We take an immersion
(possibly with double points) of the graph in $\matR^2$, in such a way
that at each vertex all the edges start with a positive
upward component
of the velocity, as in~(\ref{fig:VERTEX}) below.
Then the numbering $0,1,\dots,n$ is simply taken from left to right. Direction
and label of an edge are marked on the edge itself.

\begin{definition}
{\em A fat $n$-graph is said to be
{\em oriented} if all its edges are labeled by {\em even} permutations.
The set of all such graphs is denoted by  $\OFG{n}$.}
\end{definition}

It is now straight-forward to encode the Feynman diagram expansion
(\ref{feyn:ntensor}) of the $n$-tensor model in terms of a sum over
all {\it fat $n$-graphs}. We proceed as follows. We first label all the
initial portions of edges by a multi-index $k$. We then associate
tensor expressions to vertices and edges of the fat graph, according
to the following rules:
\begin{eqnarray}\nonumber 
   \begin{array}{c}\setlength{\unitlength}{1 pt}
   \begin{picture}(70,35)
   \put( 0, 0){\makebox(10,10){$\idx{j}$}}
   \put(20,10){\oval(30,30)[tl]}
   \put(20,25){\vector(1,0){30}}
   \put(30,27){\makebox(10,10){$\sigma$}}
   \put(50,10){\oval(30,30)[tr]}
   \put(60, 0){\makebox(10,10){$\idx{i}$}}
   \end{picture}\end{array}
   \Rightarrow  G^\idx{\tau}_{\Balpha^\idx{j}\Balpha^\idx{i}}
\qquad,\qquad
   \begin{array}{c}\setlength{\unitlength}{1 pt}
   \begin{picture}(80,35)
   \put(40,15){\oval(70,30)[b]}
   \put(40,15){\oval(40,30)[b]}
   \put(40, 0){\circle*{3}}
   \put(35,10){\makebox(10,10){$\ldots$}}
   \put( 5, 7){\makebox(10,10){$\scriptstyle 0$}}
   \put(20, 7){\makebox(10,10){$\scriptstyle 1$}}
   \put(75, 7){\makebox(10,10){$\scriptstyle n$}}
   \put( 0,20){\makebox(10,10){$\idx{i_0}$}}
   \put(15,20){\makebox(10,10){$\idx{i_1}$}}
   \put(70,20){\makebox(10,10){$\idx{i_n}$}}
   \end{picture}\end{array}
   \Rightarrow  V^{\Balpha^\idx{i_0}\ldots \Balpha^\idx{i_n}}
\end{eqnarray}
where $\tau=\sigma\circ (1\;n)$. In this way we obtain a tensor
expression in which each multi-index $\Balpha^\idx{i}$ appears
exactly twice. We denoted by $[[G]]$ the sum over all
the possible values of the multi-indices $\Balpha^\idx{i}$ of this
tensor expression.

Using the definition just given, we can now plug into (\ref{feyn:ntensor})
the explicit formula for the propagators
given in Proposition~\ref{GM:rel}, and rewrite $Z_n[N,\lambda]$
as a sum over all fat $n$-graphs. One only needs to be careful about
multiplicities. Denoting by $\nu_0(G)$ and $\nu_1(G)$ the numbers
of vertices and edges of a fat graph $G$, the right expression is computed to be
\begin{eqnarray}\label{feyn2:ntensor}
Z_n[N,\lambda] & = & 1+ \sum_{G\in\OFG{n}}\ w_n(G)\cdot \lambda^{\nu_0(G)}\cdot [[G]], \\
\nonumber 
w_n(G) & = & \frac{\mu(G)}{\nu_0(G)!\cdot (n+1)^{\nu_0(G)}\cdot (n!)^{\nu_1(G)}},
\end{eqnarray}
where $\mu(G)$
is the number of the inequivalent ways of labeling the vertices of $G$
with $\nu_0(G)$ symbols. Note also that $\nu_1(G)=(n+1)\nu_0(G)/2$, and that the factors
$\nu_1(G)!$ and $2^{\nu_1(G)}$ originally appearing in (\ref{feyn:ntensor})
get simplified during the computation.

\section{Complex associated to a fat graph}\label{fat:to:poly:section}

Given a finite set $\calS$ of (disjoint) simplices, all having
the same dimension, we call {\em face-pairing} $\calP$ on $\calS$
a set of simplicial homeomorphisms between codimension-1 faces of
the elements of $\calS$, where each such face appears exactly
once as the source or target of a homeomorphism. We interpret
$\calP$ as a set of glueing instructions between the simplices of
$\calS$, and we denote by $\calS/\calP$ the space resulting from
the glueing. We will often implicitly assume that $\calS/\calP$
is connected. We will denote by $\GS{n}$ the set of all spaces of
the form $\calS/\calP$ where all the elements of $\calS$ are
$n$-simplices. So an element of $\GS{n}$ is a simplicial complex
with a fixed ``singular'' triangulation.

We now define a map from the set $\FG{n}$ of fat $n$-graphs to
the set $\GS{n}$ of spaces obtained by glueing $n$-simplices. To
each vertex $v$ of $G\in\FG{n}$ we associate an $n$-simplex
$S(v)$ with vertices labeled $p_i(v)$, $i=0,\ldots,n$. To each
initial portion of edge at $v$ we associate an $(n-1)$-face of
$S(v)$, according to the rule
\begin{equation} \label{fig:VERTEX}
   \begin{array}{c}\setlength{\unitlength}{1 pt}
   \begin{picture}(180,50)
   \put(  0,35){\makebox(10,10){$\theta(p_0(v)$)}}
   \put( 45,35){\makebox(10,10){$\theta(p_1(v)$)}}
   \put( 90,30){\oval(160,40)[b]}
   \put( 90,30){\oval( 80,40)[b]}
   \put( 85, 0){\makebox(10,10){$v$}}
   \put( 90,10){\circle*{4}}
   \put( 90,30){\oval( 30,40)[bl]}
   \put(100,20){\makebox(10,10){$\cdots$}}
   \put(165,35){\makebox(10,10){$\theta(p_n(v)$)}}
   \end{picture}\end{array}
\end{equation}
where $\theta(p_i(v))$ represents the face opposite to $p_i(v)$.
Each edge of $G$ determines a pairing (simplicial identification)
between the $(n\!-\!1)$-faces associated to its ends,
as described now. First, to each vertex of $G$ we associate
the following object:
\begin{equation} \label{fig:VERTEX2}
   \begin{array}{c}\setlength{\unitlength}{1 pt}
   \begin{picture}(250,50)
   \put(28,15){\makebox(0,0)[bl]{$\theta(p_0(v)$)}}
   \put( 5,30){\line( 1, 0){40}}
   \put( 5,35){\line( 0,-1){ 5}}
   \put( 0,37){\makebox(10,10){$\scriptstyle i_1^0$}}
   \put(15,35){\line( 0,-1){ 5}}
   \put(10,37){\makebox(10,10){$\scriptstyle i_2^0$}}
   \put(25,37){\makebox(10,10){$\ldots$}}
   \put(45,35){\line( 0,-1){ 5}}
   \put(40,37){\makebox(10,10){$\scriptstyle i_n^0$}}
   \put(98,15){\makebox(0,0)[bl]{$\theta(p_1(v)$)}}
   \put(75,30){\line( 1, 0){40}}
   \put(75,35){\line( 0,-1){ 5}}
   \put(70,37){\makebox(10,10){$\scriptstyle i_1^1$}}
   \put(85,35){\line( 0,-1){ 5}}
   \put(80,37){\makebox(10,10){$\scriptstyle i_2^1$}}
   \put(95,37){\makebox(10,10){$\ldots$}}
   \put(115,35){\line( 0,-1){ 5}}
   \put(110,37){\makebox(10,10){$\scriptstyle i_n^1$}}
   \put(228,15){\makebox(0,0)[bl]{$\theta(p_n(v)$)}}
   \put(205,30){\line( 1, 0){40}}
   \put(205,35){\line( 0,-1){ 5}}
   \put(200,37){\makebox(10,10){$\scriptstyle i_1^n$}}
   \put(215,35){\line( 0,-1){ 5}}
   \put(210,37){\makebox(10,10){$\scriptstyle i_2^n$}}
   \put(225,37){\makebox(10,10){$\ldots$}}
   \put(245,35){\line( 0,-1){ 5}}
   \put(240,37){\makebox(10,10){$\scriptstyle i_n^n$}}
   \put(125,0){\circle*{4}}
   \put(125,30){\oval(200,60)[b]}
   \put(125,30){\oval(60,60)[b]}
   \put(125,30){\oval(100,60)[br]}
   \put(160,20){$\cdots$}
   \end{picture}\end{array}
\end{equation}
where the sequence $(i_1^k,i_2^k, \cdots ,i_n^k)$ depends on whether
$n\cdot k$ is even or odd. If $n\cdot k$ is even then the sequence is
$(k-1,k-2,\cdots,k-n)$, with indices meant modulo $n+1$, while if
$n\cdot k$ is odd then the sequence is $(k+1,k+2,\cdots,k+n)$, with
indices again modulo $n+1$. Now an edge of $G$ can be pictured
as follows:
\begin{equation} \label{fig:GLUING}
   \begin{array}{c}\setlength{\unitlength}{1 pt}
   \begin{picture}(200,60)
   \put(25, 5){\line( 0, 1){10}}
   \put(27, 0){$\theta(p_{j_0}\!(v))$}
   \put( 5,15){\line( 1, 0){40}}
   \put( 5,20){\line( 0,-1){ 5}}
   \put( 0,22){\makebox(10,10){$\scriptstyle j_1$}}
   \put(15,20){\line( 0,-1){ 5}}
   \put(10,22){\makebox(10,10){$\scriptstyle j_2$}}
   \put(25,22){\makebox(10,10){$\ldots$}}
   \put(45,20){\line( 0,-1){ 5}}
   \put(40,22){\makebox(10,10){$\scriptstyle j_n$}}
   \put( 40,35){\oval(30,30)[tl]}
   \put( 40,50){\vector(1,0){70}}
   \put( 70,52){\makebox(10,10){$\sigma$}}
   \put(110,35){\oval(30,30)[tr]}
   \put(125, 5){\line( 0, 1){10}}
   \put(127, 0){$\theta(p_{i_0}\!(w))$}
   \put(105,15){\line( 1, 0){40}}
   \put(105,20){\line( 0,-1){ 5}}
   \put(100,22){\makebox(10,10){$\scriptstyle i_1$}}
   \put(115,20){\line( 0,-1){ 5}}
   \put(110,22){\makebox(10,10){$\scriptstyle i_2$}}
   \put(125,22){\makebox(10,10){$\ldots$}}
   \put(145,20){\line( 0,-1){ 5}}
   \put(140,22){\makebox(10,10){$\scriptstyle i_n$}}
   \end{picture}\end{array}
\end{equation}
and we associate to the edge the map from $\theta(p_{i_0}(v))$ to
$\theta(p_{j_0}(w))$ which maps $p_{i_k}(v)$ to
$p_{j_{\tau(k)}}(w)$ where $\tau= \sigma\circ (1\;n)$. Summing
up, we have associated to $G\in\FG{n}$ a set $\calS$ of
$n$-simplices and a face-pairing $\calP$ on this set. The result
is then a triangulated complex $\calS/\calP$, so indeed a map
$\FG{n}\to\GS{n}$ has been defined.

\begin{rem}\label{all:induced:ori:rem}
{\em If we give to $S(v)$ the orientation induced
  by the ordering $p_{0}(v)$, $p_{1}(v)$, \ldots, $p_{n}(v)$,
  and to its faces the orientation as portions of the boundary, then
  the ordering chosen in (\ref{fig:VERTEX2}) is always a positive one.
  In particular, a face-glueing  as in (\ref{fig:GLUING}) reverses
  the induced orientation precisely when $\sigma\in \permu_n$ is an even
  permutation. This explains Fig.~\ref{TM3dim} and answers the
  naturality issue raised in Remark~\ref{weird:3D:choice}.}
\end{rem}

\begin{rem}\label{circuit:rem}
{\em Taking Fig.~\ref{TM3dim} and Fig.~\ref{TM4dim} as models,
one can rather easily turn (\ref{fig:VERTEX2}) and
(\ref{fig:GLUING}) into rules which allow to associate to a fat
graph $G$ a pattern of circuits ({\em i.e.} maps from the circle)
on the graph. If there are $\nu_2(G)$ of these circuits and we
attach $\nu_2(G)$ discs $D^2$ to $G$ along them, we get a
polyhedron with precisely three types of points. Namely, the
neighbourhood of a point is either a plane, or the union of $n$
half-planes with common boundary line, or the infinite cone over
the 1-skeleton of an $n$-simplex. It is not hard to see that this
polyhedron is actually the 2-skeleton of the cellularization dual
to the triangulation defined by the graph as just explained. We
can then interpret the fat graph  as a way of describing the dual
2-skeleton of a triangulation. In particular, this dual 2-skeleton
determines the triangulation itself. Note that, in general, a
simplicial complex is {\em not} determined by the 2-skeleton dual
to the decomposition into simplices. This is however true if the
complex is obtained by glueing codimension-1 faces of simplices,
as in the case of complexes defined by fat graphs. }
\end{rem}

\begin{rem}{\em For a triangulation of a manifold it is always
automatically true that it is obtained by glueing codimension-1
faces of simplices. Moreover, if the manifold has empty boundary,
any codimension-1 face belongs to precisely two simplices (or to
only one but with multiplicity two). So the triangulation defines
an element of $\GS{n}$, and in particular it comes from a fat
graph.}
\end{rem}

\begin{rem}\label{equiv:rel:rem}
{\em As already mentioned, the geometric construction described in
this section can be used to define a map $\FG{n}\to\GS{n}$. Any
space presented as $\calS/\calP$ arises from some fat graph, so
the map is surjective. Moreover, given an element of $\GS{n}$, if
we choose a direction for the glueings and a numbering from $0$
to $n$ of the vertices of each simplex, a unique fat graph is
determined. This shows that the map $\FG{n}\to\GS{n}$ is a
bijection, provided one defines on $\FG{n}$ the equivalence
relation which corresponds to the arbitrariness of the choices
just described. In other words, up to taking the appropriate
identifications, a fat graph and a space obtained by glueing
simplices are one and the same thing.}
\end{rem}

\paragraph{Equivalence of fat graphs} We describe here how
to turn the map $\FG{n}\to\GS{n}$ into a bijection. Let us first spell out
the equivalence relation implicit in the definition of $\GS{n}$. Two
spaces $\calS/\calP$ and $\calS'/\calP'$ are identified if they are {\em
combinatorially equivalent}, namely if there exists a simplicial
isomorphism $\phi:\calS\to\calS'$ such that each glueing of $\calP$
corresponds via $\phi$ to one of $\calP'$ or to the inverse of one of
$\calP'$. The equivalence relation on $\FG{n}$ needed to make
$\FG{n}\to\GS{n}$ bijective is now generated by two moves. The first move
consists in reversing the direction of an edge and at the same time
replacing its colour $\sigma$ by $(1\;n)\cdot\sigma^{-1}\cdot (1\;n)$.
The second move takes place at a vertex and depends on the choice of
$\lambda\in\permu_{\{0,\dots,n\}}$, as described in
Fig.~\ref{permuvert}-left,
\figura{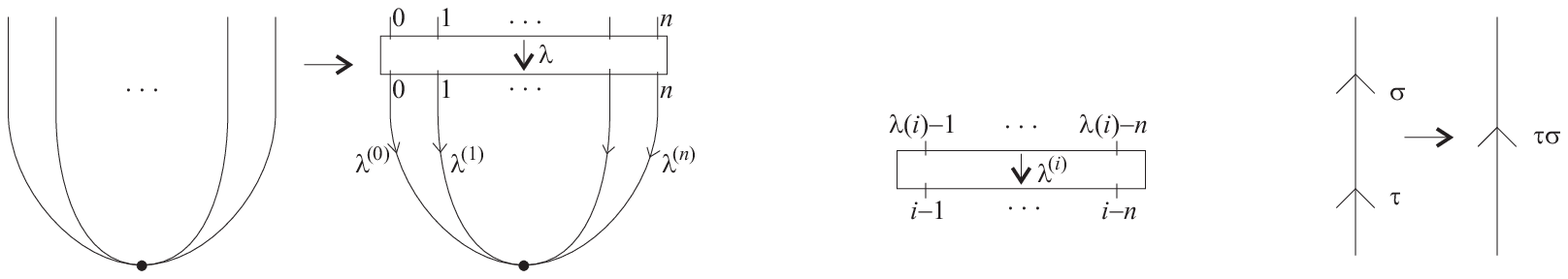}{2.5cm}{The move which
corresponds to a change of numbering of vertices of a simplex}
where as usual $\,\downarrow\!\lambda\,$ means that the $i$-th strand on
the bottom is matched with the $\lambda(i)$-th strand on the top. Here
$\lambda^{(i)}:\{0,\ldots,n\}\setminus\{i\}\to
\{0,\ldots,n\}\setminus\{\lambda(i)\}$ is just the restriction of $\lambda$,
and its action on strands is defined as above, as suggested in
Fig.~\ref{permuvert}-centre. For the sake of simplicity in
Fig.~\ref{permuvert}-centre we have assumed that both $n\cdot i$ and $n\cdot\lambda(i)$
are even; in general the rule described after (\ref{fig:VERTEX2})
should be employed.
Note that after the second move some edges can have multiple colours. To get rid of
them, one should employ the (very natural) association rule
described in Fig.~\ref{permuvert}-right
(but note that to apply the rule one may first need to reverse the
direction of the edge, using the first move).

\paragraph{General vertex function}

We can now provide the general definition of the vertex function to be used in
(\ref{Z:ntensor}), so that it translates precisely
(\ref{fig:VERTEX2}) under the correspondence between
fat graphs and tensor expressions defined in
(\ref{eq:GR}) and (\ref{eq:GRbis}). Namely, we define
\begin{equation} \label{EQ:genVertex}
V^{\Balpha^\idx{0}\ldots \Balpha^\idx{n}}
   = \prod_{r<s} \delta^{\alpha^\idx{r}_{m(r,s)} \alpha^\idx{s}_{n(r,s)}}
\end{equation}
where $m(r,s)$ and $n(r,s)$ are implicitly defined by the
condition that, using the sequence $i^k_l$ defined in
(\ref{fig:VERTEX2}), $i^r_{m(r,s)}=s$ and $i^s_{n(r,s)}=r$,
respectively. So for instance when $n$ is even we have that $m(r,s)$ and
$n(r,s)$ are given respectively by $r-s$ modulo $n+1$ and $s-r$ modulo $n+1$.
Of course this definition is coherent with the vertex function
already defined in the case of rank-3 and rank-4
tensor models given in Fig.~\ref{TM3dim}
and Fig.~\ref{TM4dim}, respectively.

We go back now to (\ref{Z:DT}) and (\ref{feyn2:ntensor}), in order to
compare them. First of all we note that, by the very choice of the
vertex function of (\ref{EQ:genVertex}), the evaluation of the tensor
expression corresponding to a fat $n$-graph $G$ involves the
computation of precisely $\nu_2(G)$ traces of the Kronecker $\delta$,
so $[[G]]=N^{\nu_2(G)}$.
Here $\nu_2(G)$ is the number of circuits determined on the graph $G$
by equations (\ref{fig:VERTEX2}) and (\ref{fig:GLUING}) as already
explained in Remark~\ref{circuit:rem}. In the same remark we have also
noted that $\nu_2(G)$ is precisely the number of codimension-2 faces
of the complex associated to $G$. Note that the evaluation is indeed
invariant under the moves which define the equivalence relation on
$\FG{n}$.  This shows that when $G$ defines a manifold with a
triangulation $\calT$, the evaluation $[[G]]$ which appears in
(\ref{feyn2:ntensor}) is precisely the same as the evaluation
$\exp(-k_n \nu_n(\calT) + h_{n-2} \nu_{n-2}(\calT))$ of $\calT$ which appears
in (\ref{Z:DT}), provided one chooses the constants
in such a way that $h_{n-2}=\log N$ and
$k_n=- \log(\lambda)$.

Summing up, with the choice of constants just described, we can split
the sum in (\ref{feyn2:ntensor}) according
to whether $G\in\OGS{n}$ defines a manifold or not (in the latter case we
will just write
$G\not\in{\rm Top}_n$ for the sake of simplicity). Namely:
\begin{equation} \label{Z:GMM}
\begin{split}
Z_{n}[N,\lambda] = 1
  &+ \sum_{M\in {\rm Top}_n^{+}}\ \sum_{\calT\in {\rm Tria}(M)}
  w_n(\calT)\;{\rm e}^{-k_n \nu_n(\calT) + h_{n-2} \nu_{n-2}(\calT)}
  \\
  &+ \sum_{G\in\OGS{n},\ G\not\in {\rm Top}_n}\
w_n(G)\cdot \lambda^{\nu_0(G)}\cdot N^{\nu_2(G)}.
\end{split}
\end{equation}
where the weight of $\calT$ is computed by picking any $G\in\OFG{n}$ which defines $\calT$,
and setting $w_n(\calT)$ equal to $w_n(G)$ times the number of all different such $G$'s.
This number is of course just equal to the number of different elements of
$\OFG{n}$ which are equivalent to $G$ under the moves defined in the previous paragraph.

Equation (\ref{Z:GMM}) shows that, to understand the exact
relation between the partition function of the $n$-tensor model
and that of the Dynamical Triangulation, one needs to be able to
tell which elements of $\OGS{n}$ define $n$-manifolds. This is the
topic of next section, where we deal with the more general case of
$\GS{n}$, which would arise anyway by restricting to a real
integration variable in the partition function~(\ref{Z:ntensor}).

\section{Simplicial glueings which define manifolds}\label{main:section}
The discussion of the preceding sections motivates a purely topological
question, to state which we set up notations which will be used throughout
this section.  Fix an integer $n$ and consider a finite number of
copies $\calS=\{\Delta_1,\dots,\Delta_k\}$ of the standard $n$-simplex.
Let $\calP$ be a face-pairing on $\calS$, denote $\calS/\calP$ by $X$ and
let $\pi:\calS\to X$ be the natural projection. Let $X^*$ be the space
obtained from $X$ by removing the projection of the vertices of the
$\Delta_i$'s.

\begin{que}\label{X:question}
When is $X$ a (closed) $n$-manifold?
\end{que}

\begin{que}\label{X^*:question}
When is $X^*$ an (open) $n$-manifold?
\end{que}

A remark is in order. The reason why we consider also $X^*$ and not $X$
only is that a satisfactory answer exists for $X$ only if $n\leq 3$, while
if $n=4$ we can provide such an answer for $X^*$ but not for $X$. Moreover
the answer for $X^*$ and $n=3$ is very easily expressed.

\paragraph{PL category} Before proceeding we need to be more specific
about the category in which we ask our questions. Since we are dealing
with simplices, the obvious category to use is the piecewise-linear one
(PL for short). All the definitions and results we will mention about PL
topology may be found in~\cite{ro:sa}

The space $X$ has an obvious (finite) PL structure.  Note however that the
projections of the $\Delta_i$'s {\em do not} provide in general a
triangulation of $X$, because the restriction of $\pi$ to $\Delta_i$ may
well be non-injective. However, if we triangulate each $\Delta_i$ using a
fine enough subdivision, we do get a triangulation in the projection.
(One sees that $n$ iterations of the barycentric subdivision always
suffice, but we do not insist on this point.)

The space $X^*$ also has a PL structure, obtained by choosing
(infinite but locally finite)  triangulations of the
$\Delta_i^*=\Delta_i\setminus\{{\rm vertices}\}$, in a way which
is consistent with the glueings. Rather than providing the
details of this construction, we show how to realize $X^*$ as a
subset of another polyhedron $X^\partial$, which will allow us to
understand $X^*$ better. Let us consider in $\Delta_i$ the second
barycentric subdivision, and let us remove the open stars of the
original vertices, thus getting a polyhedron $\Delta_i^\partial$.
(Recall that the {\em star} of a vertex in a simplicial complex
is the union of all the simplices containing the vertex; for the
{\em open star} only the interior of these simplices is taken.)
Since the elements of $\calP$ are simplicial maps, they restrict
to glueings between the $\Delta_i^\partial$'s.  We denote by
$X^\partial$ space resulting from these glueings. Of course
$X^\partial$ has a (finite) PL structure. It easily follows that
$X^*$ embeds in $X^\partial$ as an open subset (see
Remark~\ref{X*:in:Xpartial} below). The natural question to ask
about $X^\partial$ is now:

\begin{que}\label{X^partial:question}
When is $X^\partial$ a (compact) $n$-manifold (with boundary)?
\end{que}

Another natural question is:
\begin{que}\label{orient:que}
Provided $X$ (or $X^\partial$) is a manifold, when is it
orientable?
\end{que}

The answer to this question is actually easy. Knowing that a common
codimension-1 face is induced opposite orientations from two adjacent
simplices, we are led to the following condition on $(\calS,\calP)$ and
the subsequent straight-forward result:
\begin{enumerate} \item[\Ori] Up
to reversing the natural orientation of some of the $\Delta_i$'s, all
face-pairings in $\calP$ reverse the induced orientation.
\end{enumerate}
(\Ori\ stands here for `\Ori entability of the manifold.')

\begin{prop}\label{orient:prop}
Let $X$ and $X^\partial$ be manifolds. They are orientable if and only
if \Ori\ holds.
\end{prop}

To face the other questions raised we will need to recall what a PL
$n$-manifold exactly is, but we first state the result which clarifies
the mutual relations. We will give a proof at the end of this section,
after having acquainted the reader to the basic techniques of PL
topology.

\begin{prop}\label{questions:relations}
\begin{enumerate}
\item\label{bounded:point} $X^*$ is an open $n$-manifold if and only if
$X^\partial$ is a compact $n$-manifold with boundary, and in this case
$X^*$ is homeomorphic to $X^\partial\setminus\partial X^\partial$.
\item\label{closed:point} $X$ is a closed $n$-manifold if and only if
$X^\partial$ is a compact $n$-manifold with boundary and all the
components of $\partial X^\partial$ are homeomorphic to $S^{n-1}$.
\end{enumerate}
\end{prop}

Using this proposition we will only focus henceforth on $X$ and $X^*$,
leaving $X^\partial$ in the background.

So, let us start with some basics of PL topology. Let $X$ be a polyhedron
with a triangulation $\calK$, and let $p\in X$. Assuming first $p$ is a
vertex of $\calK$, we define its {\em link} as
$$\lk_\calK(p)=\bigcup\Big\{\sigma\in\calK:
p\not\in\sigma,\ \exists\tau\in\calK\ {\rm s.t.}\
\tau\supset\{p\}\cup\sigma\Big\}.$$
If $p$ is not a vertex we consider a subdivision of $\calK$ in which $p$
is a vertex, and consider its link there. It is a fact that the link is
independent of $\calK$ up to PL homeomorphism, so we will just write
$\lk_X(p)$. Some examples of links in a small 2-dimensional polyhedron are
shown in Fig.~\ref{links}.
\figura{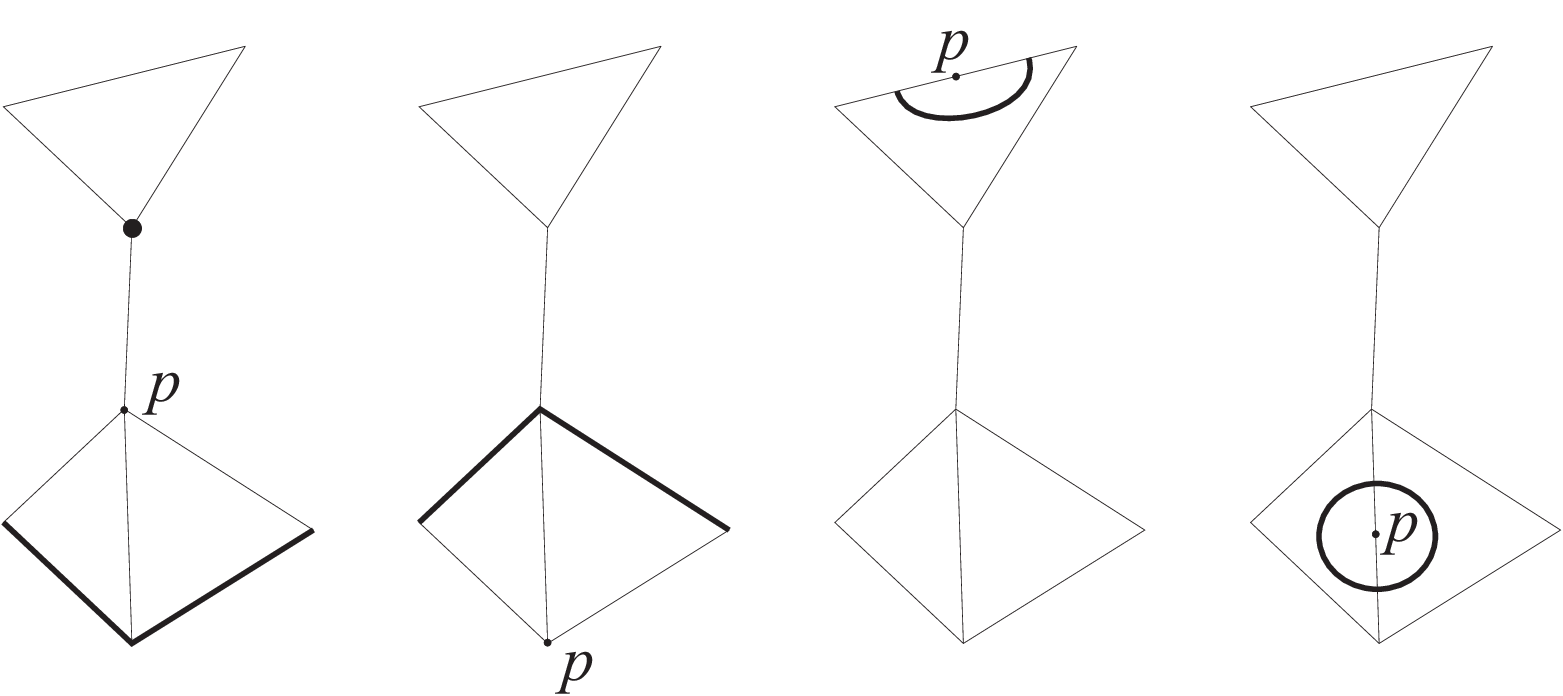}{4cm}{The links of the various
points $p$ are shown in bold.}
Note that the star of a point $p$,
mentioned above, is just the cone from $p$ over the link of $p$. This
remark motivates the following fact
(which is actually often used as a definition): {\em a
polyhedron is a PL $n$-manifold if and only if the link of every point is
homeomorphic either to the $(n-1)$-sphere $S^{n-1}$ (for interior points)
or to the closed $(n-1)$-disc $D^{n-1}$ (for boundary points).}

\begin{rem}\label{X*:in:Xpartial}{\em Even when $X^\partial$ is
not a manifold we can define $\partial X^\partial$ as the
projection of the links of the vertices of $\calS$ (in the second
barycentric subdivision). Then we always have
$X^*=X^\partial\setminus\partial X^\partial$.}
\end{rem}

\paragraph{Dimension 2} When the dimension is 2,
Question~\ref{X:question} always has a positive answer. To see this, note
that (in all dimensions) the link of any point $p\in X$ is obtained by
glueing the links (in the $\Delta_i$'s) of its preimages $\pi^{-1}(p)$ via
the pairings induced from those in $\calP$.  Now, for $n=2$, either
$\pi^{-1}(p)$ is a single point in the interior of a triangle, and its
link is $S^1$, or it is made of points on the boundary of the triangles,
and all the links are segments. The face-pairing $\calP$ induces the
identification in pairs of the endpoints of these segments, and the result
is again $S^1$.

Before turning to dimension 3, we make another general remark. Since the
restriction of the projection $\pi$ to each $\Delta_i$ can be far from
injective, one cannot predict in general how many points a fibre
$\pi^{-1}(p)$ will contain, but there are two exceptions. First, if $q$
belongs to the interior of one of the $\Delta_i$'s, then $q$ is not glued
to any other point, so $\pi^{-1}(\pi(q))=\{q\}$, and the link of $\pi(q)$
certainly is $S^{n-1}$. Second, if $q$ lies in the interior of a
codimension-1 face of one of the $\Delta_i$'s, then $q$ gets glued to
another point only, {\em i.e.}~$\pi^{-1}(\pi(q))=\{q,q'\}$. In this case
$\lk_X(\pi(q))$ is obtained from $\lk_{\Delta_i}(q)$ and
$\lk_{\Delta_i'}(q')$, which are both homeomorphic to $D^{n-1}$, by an
identification of their common boundary $S^{n-2}$, and the result is again
$S^{n-1}$. This shows that when facing Questions~\ref{X:question}
or~\ref{X^*:question}, one only needs to compute the links of the points
$\pi(q)$ where $q$ lies in a face of codimension 2 or more.

\paragraph{Dimension 3} To answer Question~\ref{X^*:question} for
$n=3$ we introduce now the following condition on $(\calS,\calP)$.
Note that this condition, as all
others we will introduce, can be checked algorithmically in a very easy
way once a concrete encoding of the pair $(\calS,\calP)$ is given.

\begin{enumerate}
\item[\Dir] The edges of the $\Delta_i$'s can be given an orientation so
that all the elements in $\calP$, when restricted to edges, match the orientation.
\end{enumerate}
(Here \Dir\ stands for `\Dir ection of the edges,' the term {\em
orientation} having been already taken up before.)

\begin{prop}\label{$3$-dim:prop}
$X^*$ is an open $3$-manifold if and only if \Dir\ holds.
\end{prop}

\dim{$3$-dim:prop} One sees quite easily that for all points of $X^*$
except the projections of midpoints of edges the link in $X$ is always
$S^2$, without any assumption.  Consider now an edge of one of the
$\Delta_i$'s, and let $q$ be its midpoint. If we arbitrarily choose points
$q_\pm$ lying on the same edge but on opposite sides of $q$, we see that
$\lk_{\Delta_i}(q)$ is a bigon with vertices $q_+$ and $q_-$. Considering
the glueings induced by $\calP$, we will have that the edges of the bigons
get glued in pairs. A connected space obtained by glueing in pairs the
edges of certain bigons is either $S^2$ or the projective plane, depending
on whether the two vertices of each bigon remain distinct in the glued
space or not. Now one easily sees that \Dir\ is precisely the condition
that these vertices remain distinct, and the conclusion
follows.\finedim{$3$-dim:prop}

Using the result just established and
Proposition~\ref{questions:relations}(\ref{closed:point}), we see
that in dimension 3, to check whether $X$ is a closed manifold,
we must first check \Dir, and then verify that $\partial
X^\partial$ is a union of $S^2$'s.  When we truncate the
$\Delta_i$'s to get the $\Delta^\partial_i$'s, we get triangles
on the boundary, so $(\calS,\calP)$ determines a triangulation
$\partial X^\partial$. Now, a triangulated surface is $S^2$ if
and only if it is connected and its Euler-Poincar\'e
characteristic $\chi$ is $2$. This implies that the question
whether $X$ is a manifold can be checked algorithmically.

\paragraph{Orientation in dimension 3}
We have now the following result which shows that in dimension 3 all is
really easy.  This result underlies the construction
in~\cite{manuscripta}.

\begin{prop}\label{$3$-dim:ori:prop}
If $n=3$ then \Dir\ is implied by \Ori.
\end{prop}

\dim{$3$-dim:ori:prop} We assume that all tetrahedra are oriented in such
a way that the elements of $\calP$ reverse the induced orientation.
Consider an edge $(v_0,v_1)$ belonging to a tetrahedron
$(v_0,v_1,v_2,v_3)$. Considering the triangle $(v_0,v_1,v_2)$, we can
assume that this ordering of the vertices defines the positive orientation
induced by the tetrahedron on the triangle.  Consider the face
$(v_0,v_1,v_3)$ and the face $(v_0',v_1',v_2')$ glued to it (where the
glueing respects the ordering of vertices). The assumptions easily imply
that the ordering $(v_0',v_1',v_2')$ defines the positive orientation. Now
let $v_3'$ be the other vertex of the tetrahedron which contains
$(v_0',v_1',v_2')$, and consider the face $(v_0'',v_1'',v_2'')$ glued to
$(v_0',v_1',v_3')$. Again $(v_0'',v_1'',v_2'')$ is a positive ordering.
Proceeding like this we will end up at some point with a glueing between
$(v^{(k)}_0,v^{(k)}_1,v^{(k)}_3)$ and $(v_{i_0},v_{i_1},v_2)$, with
$\{i_0,i_1\}=\{0,1\}$. Since $(v_{i_0},v_{i_1},v_2)$ is a positive
ordering, the permutation $(0,1,2)\to(i_0,i_1,2)$ is an even one, so
$i_0=0$ and $i_1=1$. Along our sequence we have considered all edges
glued to $(v_0,v_1)$, and our conclusion shows that a consistent
orientation for these edges can be chosen.\finedim{$3$-dim:ori:prop}

\paragraph{Dimension 4} Assume from now on that $n=4$. It will turn out
that $X^*$ is a $4$-manifold if and only if the same condition \Dir\
considered above, and two more conditions \Cycl\ and \Surf, are satisfied.
Rather than giving formal definitions soon, we illustrate how these
conditions arise. Recall that we have nothing to check up to codimension
1, and codimension 4 ({\em i.e.}~vertices) is ruled out of $X^*$.  So we
have codimension 2 (triangles) and 3 (edges). By analogy with dimension 3,
we will only worry at first about barycentres (later we will show that
indeed if the barycentres have spherical link then all points do).

Starting from codimension 2, let $q$ be the barycentre of a triangle $T$
contained in $\Delta_i$. The link of $q$ in $\Delta_i$ is PL homeomorphic
to $D^3$, but it is convenient to analyze its combinatorial structure.
First note that $\lk_{\Delta_i}(q)\cap T=\lk_T(q)$, so it is the boundary
of a triangle, which may be identified to $\partial T$ itself.  Now $T$ is
contained in two $3$-faces of $\Delta_i$, and $\partial\lk_{\Delta_i}(q)$
is given precisely by the intersection with these two faces. Moreover the
intersection with each one is a triangle bounded by $\partial T$. Summing
up, we may identify $\lk_{\Delta_i}(q)$ with the space shown in
Fig.~\ref{links4d} (a).
\figura{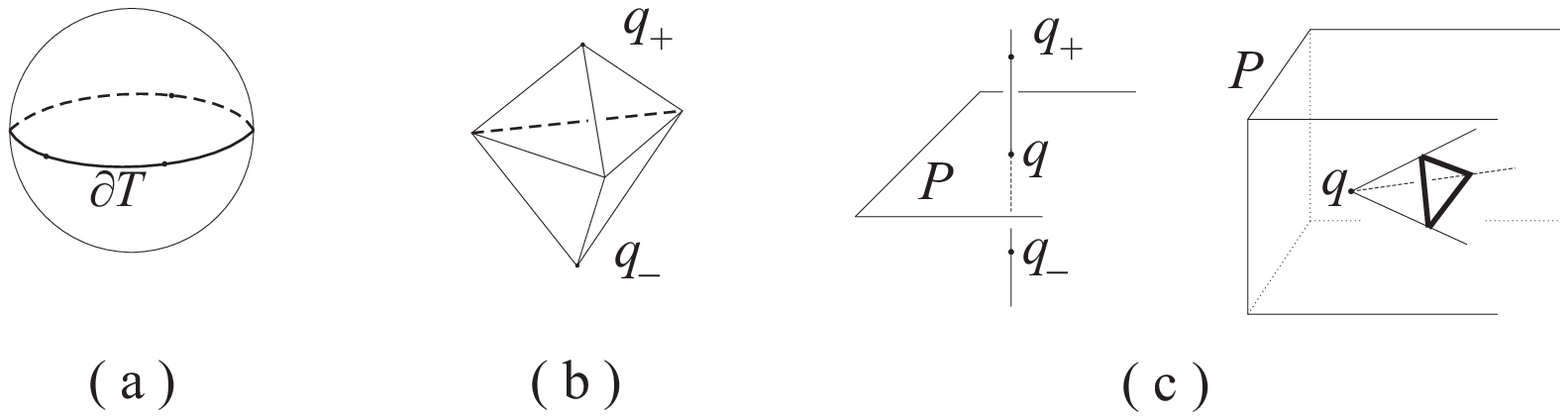}{3.5cm}{Links of the barycentres of a triangle (a) and
an edge (b). The link of the midpoint of an edge is the double cone on
the link in a cross-section (c).}
Now, to get $\lk_X(\pi(q))$ we must glue together the links of the various
points identified to $q$. Using Fig.~\ref{links4d} (a) we note that
each glueing identifies a (lower or upper) triangular hemisphere to
another one. As we proceed with the glueings, we still have a ball $D^3$
whose boundary is given by the union of two triangular hemispheres, until
the upper and lower hemisphere are glued together. Such a glueing is
determined by a permutation of the vertices of the triangle, and the
result is $S^3$ if the permutation is the identity, it is the lens space
$L_{3,1}$ if the permutation is even but non-trivial, and it is not a
$3$-manifold if the permutation is odd (of the three vertices, one is
fixed by the permutation, and it is easy to see that its link
is the projective plane).
Since the triangle glueings are precisely those
induced by the face-pairings $\calP$, we have following definition and
result:

\begin{enumerate}
\item[\Cycl] The following should happen for all triangles
$(v_0,v_1,v_2)$. Let $v_3,v_4$ be the other vertices of the same
$4$-simplex. Let $(v_0,v_1,v_2,v_4)$ be glued to
$(v_0',v_1',v_2',v_3')$, in this order. Let $v_4'$ be the other
vertex of the same $4$-simplex, and proceed until a glueing of
$(v^{(k)}_0,v^{(k)}_1,v^{(k)}_2,v^{(k)}_4)$ with
$(v_{\sigma(0)},v_{\sigma(1)},v_{\sigma(2)},v_3)$ is first found,
where $\{\sigma(0),\sigma(1),\sigma(2)\}=\{0,1,2\}$.  Then we
should have $\sigma={\rm id}$.
\end{enumerate}
(Here the name \Cycl\ comes from the fact that we look at
simplices \Cycl ically arranged around a codimension-2 face.)

\begin{prop}\label{cycl:prop}
\Cycl\ holds if and only if $\lk_X(\pi(q))\cong S^3$
for all barycentres $q$ of triangles.
\end{prop}

Note that condition \Cycl\ as stated only makes sense in dimension 4, but
one can easily devise an extension to all dimensions. In particular, the
$3$-dimensional analogue of \Cycl\ is condition \Dir\ discussed above.

We turn now to codimension 3, so we consider the midpoint $q$ of
an edge of $\Delta_i$. We arbitrarily choose two points $q_\pm$
on the same edge but on opposite sides of $q$, and we note that
$\lk_X(q)$ can be naturally identified with the `join' of
$\{q_+,q_-\}$ with $\lk_{\Delta_i\cap P}(q)$, where $P$ is a
hyperplane through $q$, orthogonal to the edge containing $q$.
The notion of `join' $A\vee B$ of two polyhedra $A$ and $B$ is
another general one from PL topology, but in the special case
where $A$ consists of two points $\{q_+,q_-\}$, the space
$\{q_+,q_-\}\vee B$ is simply the union of the two cones
$\{q_+\}\vee B$ and $\{q_-\}\vee B$ glued along the common basis
$B$.  The link of $q$ in $\Delta_i$ is shown in
Fig.~\ref{links4d} (b), and the explanation of why it can be
described like this is suggested in Fig.~\ref{links4d} (c).

Now, as above, $\lk_X(\pi(q))$ is obtained by glueing together the faces
of the various $\lk_{\Delta_{i'}}(q')$, where $\pi(q')=\pi(q)$. Each face
of $\lk_{\Delta_{i'}}(q')$ is given by $\{q'_+,q'_-\}\vee e'$, where $e'$
is an edge of the triangle $\lk_{\Delta_{i'}\cap P'}(q')$. Every glueing
between $\lk_{\Delta_{i'}}(q')=\{q'_+,q'_-\}\vee e'$ and
$\lk_{\Delta_{i''}}(q'')=\{q''_+,q''_-\}\vee e''$ maps $\{q'_+,q'_-\}$ to
$\{q''_+,q''_-\}$ and $e'$ to $e''$, and it is determined by these data.
The resulting space is then determined by the answers to the following
questions:

\begin{enumerate}
\item Can the arbitrary $q_+/q_-$ choice be made in such a way that each
$q'_+$ is glued to a $q''_+$?
\item What surface results from the triangles $\lk_{\Delta_{i'}\cap
P'}(q')$ under the edge glueings $e'\to e''$?
\end{enumerate}

Of course the answer to (1) is positive if an only if \Dir\ holds.
To answer (2)  we first formalize the construction of the surface.

\begin{enumerate}
\item[\Surf] Associate to each edge $(v_0,v_1)$ of a $4$-simplex
$(v_0,v_1,v_2,v_3,v_4)$ an abstract triangle $T(v_0,v_1)$. The edge is not
oriented, so $T(v_0,v_1)=T(v_1,v_0)$.  Denote the vertices of $T(v_0,v_1)$
by $T^{v_j}(v_0,v_1)$, for $j=2,3,4$.  For each pairing
$(v_0,v_1,v_2,v_3)\to(v'_0,v'_1,v'_2,v'_3)$, consider the edge-pairings
$$(T^{v_i}(v_j,v_k),T^{v_l}(v_j,v_k))\to
(T^{v'_i}(v'_j,v'_k),T^{v'_l}(v'_j,v'_k)),
\qquad {\rm where}\ \{i,j,k,l\}=\{0,1,2,3\}.$$
The closed surface $\Sigma(\calS,\calP)$ resulting from these
edge-pairings between the triangles should be a union of components
homeomorphic to $S^2$.
\end{enumerate}
(Here \Surf\ stands for \Surf ace.)

\begin{rem}{\rm
\begin{enumerate}
\item Each $4$-simplex determines 10 triangles, and each face-pairing
between $4$-simplices determines 6 edge-pairings between triangles.
\item Since $\Sigma(\calS,\calP)$ is intrinsically defined as a
triangulated surface, one can algorithmically check that a certain
component $\Sigma_0$ is $S^2$, by checking whether $\chi(\Sigma_0)=2$.
\end{enumerate}}
\end{rem}

\begin{prop}\label{surf:prop}
\Dir\ and \Surf\ jointly hold if and only if $\lk_X(\pi(q))\cong S^3$
for all midpoints $q$ of edges.
\end{prop}

\dim{surf:prop} If $q$ is the midpoint of $(v_0,v_1)$ then
$\lk_{\Delta_i}(q)=\{q_+,q_-\}\vee T(v_0,v_1)$, and the labels
$T^{v_i}(v_j,v_k)$ are chosen so that the glueings induced by $\calP$ are
precisely those described in \Surf. If \Dir\ and \Surf\ hold we deduce
that $\lk_X(\pi(q))\cong\{q_+,q_-\}\vee S^2\cong S^3$. Assume now that
$\lk_X(\pi(q))\cong S^3$.  A closed surface embedded in $S^3$ is
necessarily orientable, and hence transversely orientable. This implies
that we can make a consistent choice of $q_+/q_-$, so \Dir\ holds.
Therefore $S^3$ is realized as $\{q_+,q_-\}\vee \Sigma_0$ for a component
$\Sigma_0$ of $\Sigma_0(\calS,\calP)$. The link of $q_+$ in
$\{q_+,q_-\}\vee \Sigma_0$ is precisely $\Sigma_0$, so $\Sigma_0\cong
S^2$.  \finedim{surf:prop}

We can now state our main result (see below for a formal proof):

\begin{teo}\label{$4$-dim:teo}
$X^*$ is a $4$-manifold if and only if conditions \Cycl, \Dir\ and \Surf\ hold.
\end{teo}

\begin{rem}{\em This result implies that it is very easy to check
algorithmically whether $X^\partial$ is a $4$-manifold with boundary or
not. It is also easy to give a presentation of $\partial X^\partial$ as a
triangulated $3$-manifold, but the triangulation which arises is
arbitrarily complicated, so the problem of recognizing whether it is a
union of $S^3$'s is theoretically solved by the
Rubinstein-Thompson~\cite{ru:th} algorithm, but undoable in practice. This
shows that the best one can really do, using current knowledge, about
Questions~\ref{X:question}-\ref{X^partial:question}, is the closed
$3$-dimensional case and the bounded $4$-dimensional case.} \end{rem}

\paragraph{Orientation in dimension 4}
Recall that, in dimension 3, condition \Ori\ implies condition
\Dir. The situation in dimension 4 is more elaborate.

\begin{prop}\label{$4$-dim:ori:prop}
Assume $n=4$. If \Ori\ holds then the permutation $\sigma\in\permu_3$
which arises in condition \Cycl\ is automatically an even one. Moreover
\Ori\ and \Surf\ jointly imply \Dir.
\end{prop}

\dim{$4$-dim:ori:prop} The first assertion is easy: all the
triangle-pairings along the sequence in \Cycl\ reverse the induced
orientation. For the second assertion, we first note that if \Ori\ holds
then each link $\lk_{\Delta_i}(q)$ can be oriented as the boundary of the
corresponding star, and the pairings between the bigonal faces of the
$\lk_{\Delta_i}(q)$'s reverse the induced orientation. Now, if \Surf\
holds, we can choose orientations on the $T(v_0,v_1)$'s so that all
edge-pairings reverse orientation. Now we can combine the orientations of
$\lk_{\Delta_i}(q)$ and $T(v_0,v_1)$ to get a consistent choice of
$q_+/q_-$, so \Dir\ holds.\finedim{$4$-dim:ori:prop}

It is a tedious exercise, which we omit, to show that these are the only
relations which hold in general between the various conditions considered.

\paragraph{Final proofs}
The discussion accompanying the introduction of the various conditions
almost but not quite proves our main result. We complete its proof now.

\dim{$4$-dim:teo} By Propositions~\ref{cycl:prop} and~\ref{surf:prop}, if
$X^*$ is a $4$-manifold then \Cycl, \Dir\ and \Surf\ hold. To see the
opposite implication we must show that links of all points, not only of
barycentres, are homeomorphic to $S^3$.  Now condition \Cycl\ implies that
the projection $\pi$ is injective on the interior of each triangular face.
It follows that our description of $\lk_X(\pi(q))$ extends {\em verbatim}
from the centre of a triangle to any point in the interior of the same
triangle.  The same argument applies to edges, because \Dir\ implies that
$\pi$ is injective on their interior.\finedim{$4$-dim:teo}

We conclude this section with a proof omitted above. Here $n$ is again arbitrary.

\dim{questions:relations} For (1), we note that $X^*$ may be equivalently
defined by removing from $X$ not just the vertices but also their {\em
closed} stars in the second barycentric subdivision. So $X^*$ naturally
embeds in $X^\partial$, and it is clear that if $X^\partial$ is manifold
then $X^*=X^\partial\setminus\partial X^\partial$ is a manifold.

To conclude (1), we are left to show that if $X^*$ is a manifold then
$X^\partial$ is. Of course we only need to examine links of points of the
candidate boundary, {\em i.e.}~points in the projection of the bases of
the stars removed from the $\Delta_i$'s. Let $p$ be such a point and
consider $q\in\pi^{-1}(p)$. Note that $q$ lies in the link of a vertex $v$
of one of the $\Delta_i$'s.  Consider now the line in $\Delta_i$ through
$v$ and $q$, and choose on it points $q_\pm$ near $q$ so that
$q_-,q,q_+,v$ appear in this order on the line. Let $P$ be the hyperplane
in $\Delta_i$ orthogonal to the line. Then
$$\lk_{\Delta_i}(q)=\{q_+,q_-\}\vee\lk_{P\cap\Delta_i}(q),\qquad
\lk_{\Delta^\partial_i}(q)=\{q_-\}\vee\lk_{P\cap\Delta_i}(q).$$ When we
consider the glueings $\calP$, we see that all the $q_-$'s corresponding
to the various $q$'s in $\pi^{-1}(p)$ get identified to a certain point
$p_-\in X^\partial$, while the $q_+$'s get identified to a point $p_+\in
X^*\setminus X^\partial$. In the meantime the various
$\lk_{P\cap\Delta_i}(q)$'s get glued together, yielding a certain
polyhedron $W_p$. This shows that
$$\lk_{X^*}(p)=\{p_+,p_-\}\vee
W_p,\qquad\lk_{X^\partial}(p)=\{p_-\}\vee W_p.$$
Now, the assumption that $X^*$ is a manifold guarantees that
$\lk_{X^*}(p)$ is $S^{n-1}$. But the link of $p_+$ in $\{p_+,p_-\}\vee
W_p$ is $W_p$, and $S^{n-1}$ is a manifold, so $W_p$ is $S^{n-2}$.  Then
$\{p_-\}\vee W_p$ is $D^{n-1}$, and the proof of (1) is complete.

To prove (2), note that if $X$ is a manifold then $X^*$ is, because it is
an open subset of $X$. So, by (1), we see that $X^\partial$ is a manifold.
So we can assume in any case that $X^\partial$ is a manifold.  Now $X$ is
obtained from $X^\partial$ by attaching to each component of $\partial
X^\partial$ the cone based on the component, and the conclusion easily
follows.  \finedim{questions:relations}

\section{Graphic translation of conditions}\label{comb:cond:section}
We describe in this section the combinatorial counterparts in $\FG{4}$
of the conditions \Ori, \Dir, \Surf, and \Cycl\ introduced above.  We
only provide quick statements and we omit the proofs
(except for some hints concerning \Surf). One basically only needs
to plug the definitions of Section~\ref{main:section} into
the construction described in Section~\ref{fat:to:poly:section}.

Condition \Ori\ is of course just the condition that the fat graph should
be equivalent, with respect to the relation defined at the end of
Section~\ref{fat:to:poly:section}, to one of $\OFG{4}$.
An easy criterion goes as follows: \Ori\ holds if an only
we can attach a sign $\epsilon(v)$ to each vertex $v$ of the graph,
in such a way that the parity of the permutation attached to any edge
is precisely the product of the signs attached to the ends of that edge.

To translate condition \Dir, recall from~(\ref{fig:VERTEX2}) that the four
strands leaving a vertex $v$ from a branch $\theta(p_{i_0}(v))$ are
numbered by $\{0,1,2,3,4\}\setminus\{i_0\}$. Moreover a propagator matches
such a quadruple of indices with another one, belonging to the same or to
another vertex. We now pick $i_1,i_2\in\{0,1,2,3,4\}\setminus\{i_0\}$ and
arbitrarily declare that $p_{i_1}(v)\prec p_{i_2}(v)$. Note now that
$\{i_1,i_2\}$ appears precisely in three of the branches leaving $v$. We
then extend the ordering $\prec$ to the three other pairs matched to
$\{p_{i_1}(v),p_{i_2}(v)\}$ by the propagators, and we continue in a
similar way until either a contradiction to $\prec$ is reached or all
matching pairs have been visited. Condition \Dir\ si now the condition
that no contradiction is ever reached, for any initial choice of $v$ and
$i_1,i_2$.

We illustrate the procedure to check \Dir\ with an example. Consider the fat
graph of Fig.~\ref{nodir},
\figura{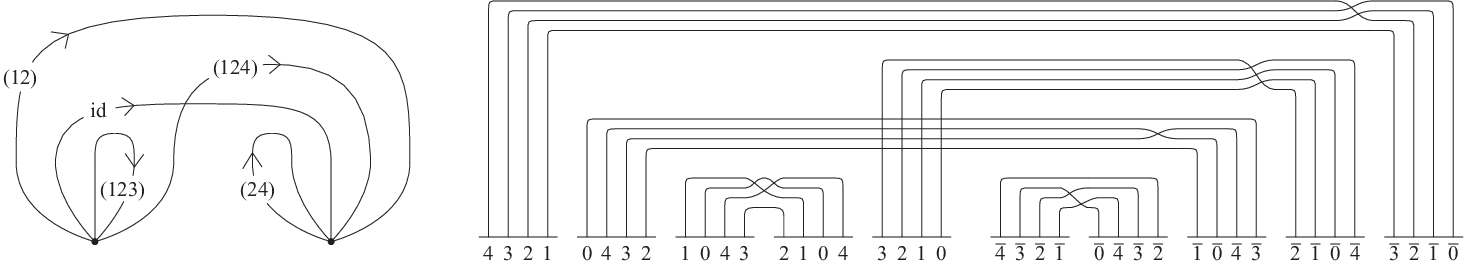}{2.8cm}{A fat graph and the corresponding propagators with
canonical labels}
let $v$ and $\bar{v}$ be its vertices, and set $i=p_i(v)$,
$\bar{i}=p_i(\bar{v})$ by simplicity. Start for instance by declaring that
$3\prec 4$. Extending this relation through the edges labeled $(1\;2)$,
id, and $(1\;2\;3)$ we respectively deduce that $\bar{0}\prec\bar{2}$,
$\bar{4}\prec\bar{0}$ and $2\prec 4$. Now for each of these pairs we have
to follow two edges (not three, because with the third one we would get
back to $3\prec 4$), and proceed so forth. Following the inverse of the
edge labeled $(2\;4)$ from $\bar{0}\prec\bar{2}$ we deduce that
$\bar{3}\prec\bar{4}$, and then following the same edge again we deduce
that $\bar{2}\prec\bar{1}$, whereas following the edge labeled $(1\;2)$
from $2\prec 4$ we deduce that $\bar{1}\prec\bar{2}$. So we have a
contradiction, and \Dir\ does not hold in this case.

Condition \Cycl\ involves the pattern of circuits already mentioned in
Remark~\ref{circuit:rem}, and obtained by joining the vertices
(Fig.~\ref{TM4dim}-centre) with the propagators (Fig.~\ref{TM4dim}-left).
To translate \Cycl\ we pick one of these circuits with an arbitrary
direction, and we follow the circuit starting from one of the
vertex-propagator junctions. We give arbitrary labels
$\alpha,\beta,\gamma$ to the three other strands at the same junction, and
we follow the labeling as we travel along the circuit, according to the
following rules. First, when we travel through a propagator, we give
matching strands the same label. Second, when we travel through a vertex,
referring to Fig.~\ref{TM4dim}-centre, we note that the various strands
come in groups of four, and we examine the relative position of the strand
at which the circuit enters the vertex. If this position is the $i$-th
one, then the circuit exits at the $(4-i)$-th position, and the labeling
rules are as follows:
$[\,|\,,\alpha,\beta,\gamma]\leftrightarrow[\alpha,\beta,\gamma,\,|\,]$ and
$[\alpha,\beta,\,|\,,\gamma]\leftrightarrow[\gamma,\,|\,,\alpha,\beta]$,
where the vertical segment $|$ represents the strand of circuit we are
following. The condition is now that, when we come back to the starting
junction, the labeling of the three other strands should be the same as at
the beginning. Some attention should be paid when a circuit travels more
than once through a junction, but the same formal rules actually apply,
one should just locally ignore that other strands are globally part of the
same circuit.

We use again the same example considered above to illustrate
the procedure for checking \Cycl. Figure~\ref{nocycl}
\figura{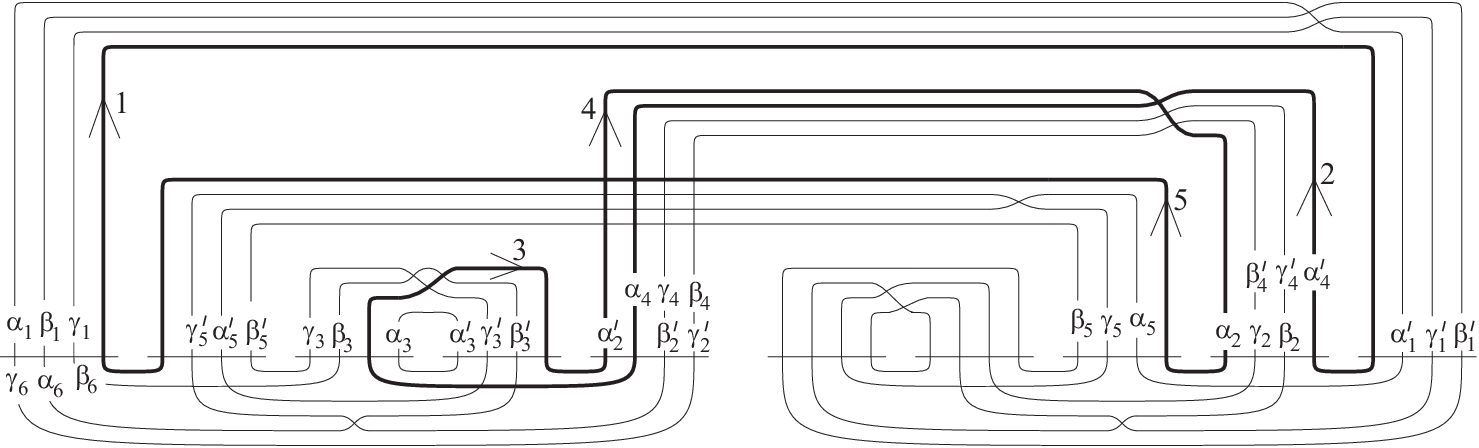}{4.2cm}{Circuits determined by a fat graph, and labels
along one of them}
shows the pattern of circuits and highlights one of them. We follow the
five edges of this circuit as suggested in the figure, starting with
labels $\alpha_1,\beta_1,\gamma_1$, denoting by
$\alpha_i',\beta_i',\gamma_i'$ the labels after the $i$-th edge, and by
$\alpha_{i+1},\beta_{i+1},\gamma_{i+1}$ the new labels after going through
a vertex. Since when we get back to the beginning of the circuit labels
are changed, \Cycl\ does not hold in this case.

We are left to translate condition \Surf. Recall
that we have to show that a certain triangulated surface $\Sigma$ is the
union of components homeomorphic to the sphere. This can be checked by
first computing the number $k$ of components, and then by checking that
$\chi(\Sigma)=2k$. To compute $k$ we will use the fact that it equals the
number of components of the 1-skeleton of the cellularization dual to the
triangulation. To check that $\chi(\Sigma)=2k$ we note that if the fat
graph has $h$ vertices, then in the triangulation of $\Sigma$ there are
$10h$ triangles and $15h$ edges, so we only need to compute the number $m$
of vertices of $\Sigma$ and check that $m=5h+2k$. Summing up, we can
express \Surf\ in the following purely combinatorial terms. Consider first
the rules of Fig.~\ref{dualskel},
\figura{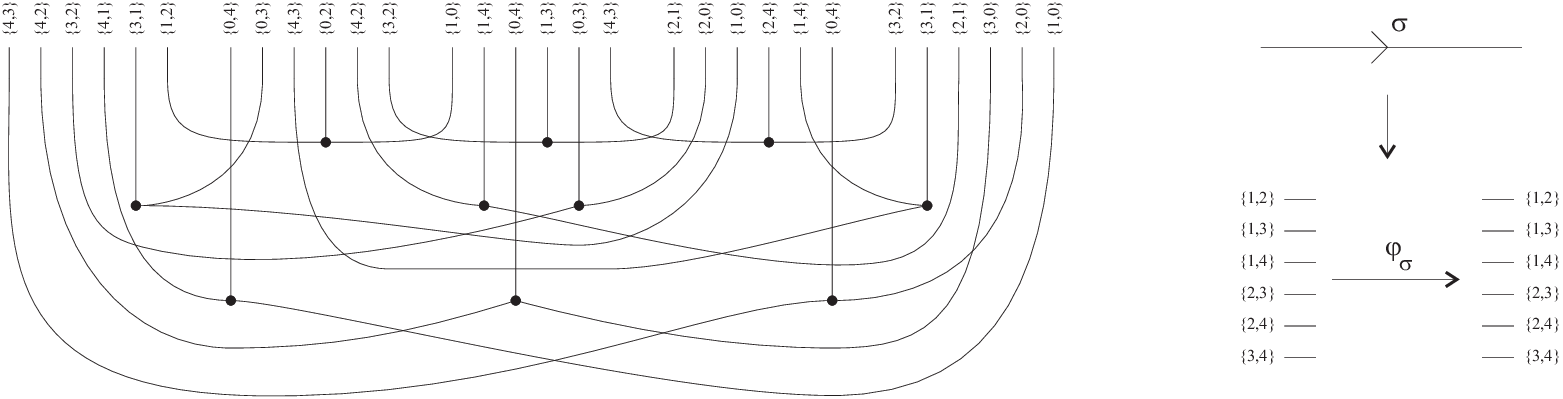}{3.8cm}{Combinatorial description of the 1-skeleton dual
to the triangulation of $\Sigma$.}
which allow to associate to the fat graph a trivalent graph. Here the rule
for the vertex is explicitly shown, and the rule for the edge is that the
strand labeled $\{i,j\}$ on the right should be matched with the strand
labeled $\varphi_\sigma(\{i,j\})$ on the left, where
$\varphi_\sigma(\{i,j\})= \{\tau(i),\tau(j)\}$, and
$\tau=\sigma\circ(1\;4)$. Denote by $k$ the number of trivalent graphs in
this picture. Next consider the rules of Fig.~\ref{vertnumb},
\figura{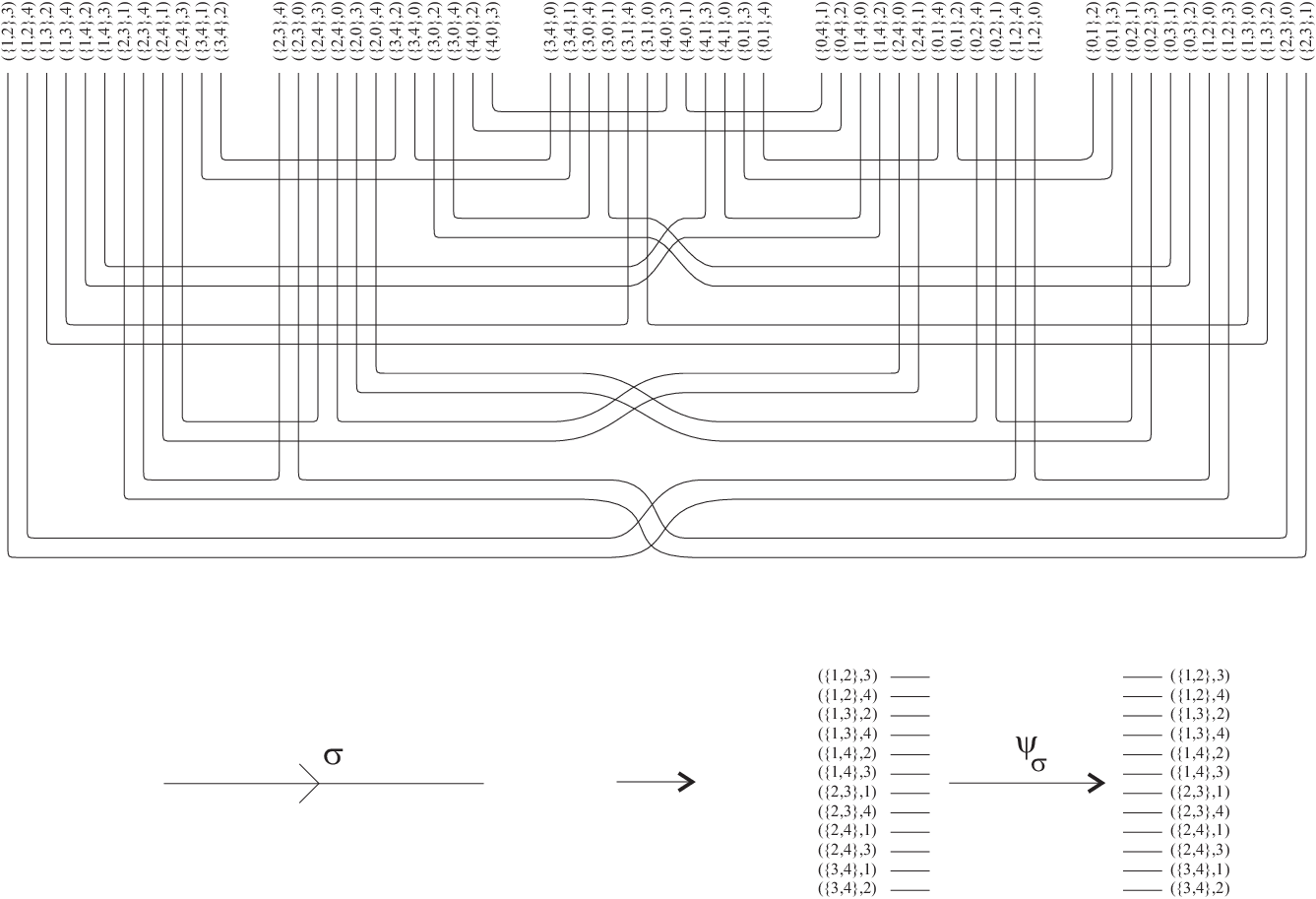}{10cm}{Circuits corresponding to the vertices of the
triangulation of $\Sigma$.}
where the meaning is as above, with
$\psi_\sigma(\{i,j\},k)=(\{\tau(i),\tau(j)\},\tau(k))$,
$\tau=\sigma\circ(1\;4)$. Denote by $m$ the number of circuits in this
figure. Then \Surf\ is precisely the condition that $m-2k$ is 5 times the
number of vertices of the fat graph.

\begin{rem}
{\em It is actually possible to attach colours in $\permu_2$ to the edges
of the trivalent graph of Fig.~\ref{dualskel}, turning it into a fat $2$-graph,
in such a way that the circuits of Fig.~\ref{vertnumb} are precisely
those defined by this graph as explained in Remark~\ref{circuit:rem}.
We have refrained from doing this and preferred to give an explicit rule.}
\end{rem}

\vskip1cm
\centerline{--------------------------------}
\vskip1cm

We thank Mauro Carfora, Annalisa Marzuoli, Carlo Rovelli and
Gianni Cicuta  for discussions and correspondence.

\end{document}